\documentclass[sigconf, nonacm]{acmart}

\usepackage{pvldb}

\usepackage[linesnumbered,ruled,vlined]{algorithm2e}
\usepackage{pdfpages}
\usepackage{graphicx}
\usepackage{enumitem}
\usepackage{makecell}
\usepackage{balance}
\usepackage{subfigure}
\usepackage{subcaption}
\usepackage{float}
\usepackage{tabularx}
\usepackage{booktabs}
\usepackage{cleveref}
\usepackage{multirow}
\usepackage{capt-of}

\usepackage{amsmath}
\usepackage{amsthm}
\usepackage{amsfonts}
\usepackage{balance}
\usepackage{dsfont}

\newtheorem{example}{Example}
\newtheorem{proposition}{Proposition}
\newcommand{\stitle}[1]{\noindent{\bf #1}}
\newcommand{\OASIS}{\textsc{Oasis}\xspace} 
\newcommand{\add}[1]{\textcolor{black}{#1}}

\crefformat{section}{\S#2#1#3} 

\renewcommand\vldbdoi{XX.XX/XXX.XX}
\renewcommand\vldbpages{XXX-XXX}
\renewcommand\vldbavailabilityurl{https://github.com/CrankyLover/OASIS} 

\author{Ziqi Zou}
\affiliation{%
  \institution{Beijing Institute of Technology}
  \city{Bejing}
  \country{China}
}
\email{ziqizou@bit.edu.cn}

\author{Zhenyan Nie}
\affiliation{%
  \institution{Beijing Institute of Technology}
  \city{Bejing}
  \country{China}
}
\email{doctorblinkwho@gmail.com}

\author{Kangfei Zhao}
\affiliation{%
  \institution{Beijing Institute of Technology}
    \city{Bejing}
  \country{China}
}
\email{zkf1105@gmail.com}

\author{Zhiwei Zhang}
\affiliation{%
  \institution{Beijing Institute of Technology}
    \city{Bejing}
  \country{China}
}
\email{zwzhang@bit.edu.cn}

\author{Aoqian Zhang}
\affiliation{%
  \institution{Beijing Institute of Technology}
    \city{Bejing}
  \country{China}
}
\email{aoqian.zhang@bit.edu.cn}

\author{Peng Wu}
\author{Xuefeng Zhao}
\affiliation{%
  \institution{China Mobile Jiutian Research}
    \city{Bejing}
  \country{China}
}
\email{wupengit@chinamobile.com}
\email{zhaoxuefengit@chinamobile.com}

\begin{document}
\title{Towards Online Adaptive LLM-based Semantic Filtering}

\begin{abstract}
Large language models (LLMs) enable semantic filtering over unstructured documents using natural language predicates, but invoking an oracle LLM on every document incurs prohibitive monetary cost and online latency. Existing model-cascade approaches reduce this cost by using a lightweight proxy model to score each document and forwarding only uncertain documents to the oracle. However, their proxy models have limited generalization capability, and they typically adopt fixed sampling budgets for proxy training and threshold calibration, which insufficiently adapts to diverse online queries. To address this, we present \OASIS, an online adaptive framework for LLM-based semantic filtering. \OASIS combines a query-agnostic document similarity prior with query-adaptive online learning and sampling. In the offline phase, \OASIS utilizes query-agnostic document similarity as a prior and constructs a document navigating spread-out graph (NSG) over document embeddings. For each online query, \OASIS trains a query-conditioned graph attention network over the document NSG to obtain proxy scores through semi-supervised node classification. To allocate labels effectively, \OASIS adopts density-aware active learning to explore uncertain score regions, selectively annotate informative documents, retrain the proxy, and stop early when additional labels are unlikely to improve the cascade. During threshold calibration, \OASIS progressively samples an adaptively sized calibration set, adapting calibration cost to the query’s score distribution. Extensive experiments demonstrate that \OASIS achieves a $1.75\times$--$5.42\times$ reduction in online latency and reduces the number of LLM-processed documents by 35\%--75\% relative to the state-of-the-art methods at an achieved accuracy of 0.9.

\end{abstract}
\vspace{-10pt}

\maketitle

\vspace{-5pt}

\section{Introduction}

Large language models (LLMs) have enabled semantic operators specified in natural language to process collections of unstructured documents~\cite{uqe, shankar2025docetl, quest, lotus, abacus, sembench}.
Unlike traditional keyword predicates or manually designed classifiers, LLMs offer strong zero-shot reasoning capability and can directly determine whether a document satisfies a complex semantic condition. Recent systems such as Lotus \cite{lotus} and Palimpzest \cite{palimpzest} exploit these  capabilities through declarative operators, allowing users to express data-processing logic over unstructured documents in  database systems. Among these operators, semantic filter is a fundamental operator.
Given a document collection and a natural language predicate $p$, a semantic filter returns the subset of documents that satisfy $p$, as judged by an LLM. This operator supports a broad range of applications, such as filtering scientific papers that claim research improvements~\cite{lotus}, identifying police records containing evidence of misconduct~\cite{shankar2025docetl}, and retrieving legal cases involving financial disputes~\cite{quest}.

\begin{figure}[t]
    \centering
    \includegraphics[scale=0.27]{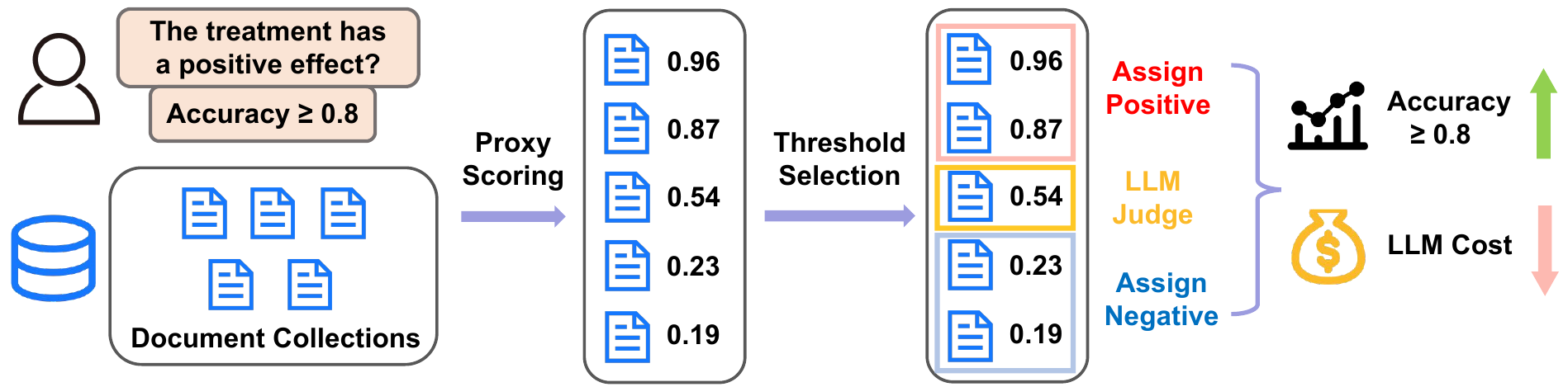}
   
    \caption{Two-stage model-cascades semantic filter paradigm.}
    \label{fig:pipeline}
\end{figure}

However, processing every document with an LLM oracle is prohibitively expensive, incurring high monetary cost and online latency. Consequently, this approach does not scale to large document collections or online query workloads.
To reduce cost and latency, existing approaches replace exact, oracle-equivalent execution with a proxy–oracle cascade, aiming to minimize the number of LLM invocations while satisfying a user-specified target accuracy $\alpha$. 
Fig.~\ref{fig:pipeline} illustrates this two-stage model-cascade paradigm. 
In the first stage, a cheaper proxy model assigns each document a proxy score indicating the likelihood that it satisfies the predicate. 
In the second stage, the system selects a threshold interval $[l,r]$: documents with scores below $l$ or above $r$ are labeled by the proxy model, whereas those with scores within $[l,r]$ are considered ambiguous and forwarded to the LLM oracle. 
For example, Lotus~\cite{lotus} and BARGAIN~\cite{bargain} use the logits of a smaller LLM as proxy scores, while ScaleDoc \cite{scaledoc} labels a subset of documents and trains a query-aware Multilayer Perceptron to predict proxy scores. For threshold calibration, Lotus learns thresholds from online samples, whereas ScaleDoc selects thresholds by reconstructing positive and negative score distributions through calibration sampling. Despite their effectiveness, these methods still face two important limitations.

First, existing proxy models provide limited generalization across online semantic filters, leading to less effective cascades. 
Under the model-cascade paradigm, documents falling into the ambiguous interval $[l,r]$ are processed by the LLM oracle, whose execution dominates the overall latency and cost. Therefore, it is worthwhile to establish a more capable and reliable proxy model if it can produce better-separated score distribution and thereby narrow the ambiguous interval. 
Ideally, negative documents should receive scores near $0$ and positive documents scores near $1$, yielding a clear bimodal distribution. However, the logits of small LLMs or the predictions of simple MLPs are not sufficiently separated for complex queries. Small LLMs may lack the reasoning capabilities required for fine-grained semantic evaluation, while MLP-based proxies rely on sufficient labeled training documents. 
In addition, both types of proxy operate on individual document representations, neglecting the semantic similarity among documents.
As a result, many documents receive ambiguous scores and have to  be forwarded to the LLM oracle, increasing both cost and online latency.

\begin{figure}[t]
    \centering
    \includegraphics[scale=0.7]{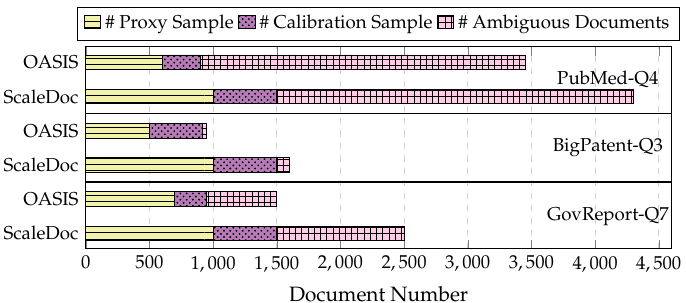}
    
    \caption{The breakdown comparison of \# LLM processing documents with ScaleDoc.}
    \label{fig:problem}
\end{figure}

Second, fixed sampling budgets fail to adapt to the difficulty of ad hoc queries. Online semantic filter queries can vary substantially in query length, selectivity, wording, and semantic complexity. For an easy query, a small number of oracle-labeled samples may  be sufficient to train a reliable proxy model and reconstruct a stable score-label distribution for threshold calibration. In such cases, a large fixed sampling budget wastes LLM calls. 
In contrast, a difficult query may require more labeled samples to obtain stable proxy scores and reliable thresholds. Insufficient training or calibration samples will introduce large bias for a proxy model and threshold calibration. 
Existing methods typically use fixed-size sampling at the two stages, i.e., a fixed training set for proxy learning or a fixed calibration set for threshold selection. This one-size-fits-all strategy cannot adapt the labeling budget to specific online queries, and finally degrade the overall two-stage framework. 

These limitations raise two challenges. First, \emph{how can we improve the effectiveness of the proxy model with only a limited number of annotated documents?} A simple proxy model is efficient but may produce poorly separated scores, whereas training a much more complex model online will incur excessive computational overhead and may overfit sparse labels. The proxy model should  therefore exploit additional prior knowledge from the unlabeled document collection while remaining efficient deployment for online execution. Second, \emph{how can we control and allocate the labeling budget for training and calibration when online queries arrive?} The difficulty lies in the system should make online decisions regarding when additional LLM annotations are necessary, which documents are informative for training, and how many samples are sufficient for calibration, in an automated and adaptive fashion.

To address these challenges, we propose \OASIS, an \underline{O}nline \underline{A}daptive LLM-based \underline{S}emantic f\underline{I}ltering \underline{S}ystem. \OASIS combines a query-agnostic document similarity prior with query-adaptive online learning and sampling. Fig. \ref{fig:problem} shows that \OASIS utilizes less labeling samples but produces more well-separated proxy score, thereby reducing ambiguous documents. To address the first challenge, \OASIS utilizes document similarity as a prior and constructs a document Navigating Spread-out Graph (NSG) in the offline phase. The graph is built over document embeddings and preserves important semantic neighborhood relationships while pruning redundant edges, yielding a sparse structure suitable for efficient online GNN training. When an online query arrives, \OASIS trains a query-conditioned Graph Attention Network over the document NSG and formulates proxy scoring as a semi-supervised node classification problem. The sampled documents annotated by LLM oracle serve as anchor nodes, and the GAT propagates their semi-supervision signals through the document NSG. This design allows \OASIS to exploit the clustering structure of semantically similar documents and produce more separated proxy scores.

To address the second challenge, \OASIS adaptively controls the LLM labeling budget in both proxy training and threshold calibration. During proxy training, \OASIS adopts an active learning strategy that iteratively selects informative documents for LLM annotation. Instead of sampling around a fixed score threshold, \OASIS detects an uncertain interval from the density shape of the current proxy score distribution. \OASIS then performs stratified sampling within this uncertain interval and retrains the proxy model. To avoid unnecessary sampling, \OASIS further introduces two early-stopping strategies based on the uncertain document ratio and the convergence of the score distribution. During threshold calibration, \OASIS progressively samples the calibration set and monitors the stability of the selected threshold pair. Easy queries can stop with fewer samples, while harder queries automatically receive more samples until the thresholds converge.

In summary, this paper makes the following contributions.
\begin{itemize}[leftmargin=*]
    \item We propose \OASIS, an online adaptive framework for LLM-based semantic filtering over large document collections.
    \item We utilize document similarity as a prior and design a query-conditioned GNN-based proxy model that combines an offline document NSG with online semi-supervised learning to improve the effectiveness of proxy model.
    \item We develop an active learning algorithm with density-aware uncertain-interval detection and early-stopping to adaptively allocate training labels. We also propose progressive calibration sampling strategy to adapt the threshold selection budget to different query difficulties.
    \item We conduct extensive experiments to show that \OASIS achieves $1.75\times$--$5.42\times$ lower online latency and reduces LLM invocation by 35\%--75\% relative to the state-of-the-art methods at an achieved accuracy of 0.9.
\end{itemize}


\section{Related Work}
\label{sec:relatedwork}

\begin{figure*}[ht]
    \centering
    \includegraphics[scale=0.45]{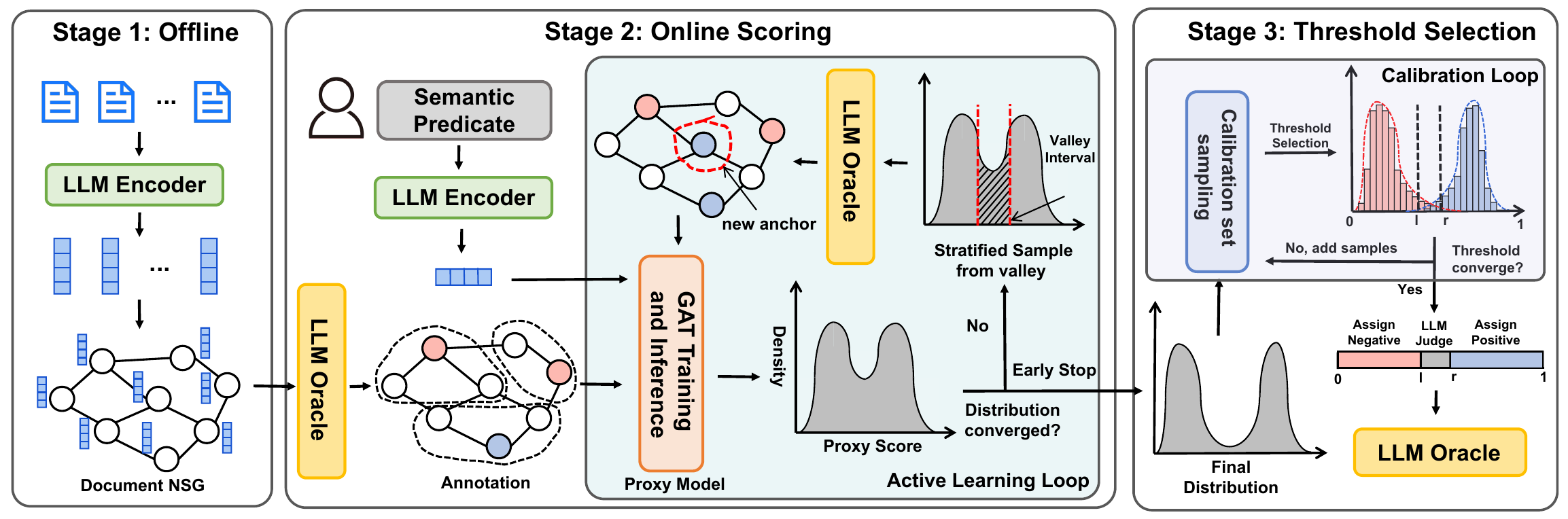}
   
    \caption{The overview of \OASIS.}
    \label{fig:overview}
\end{figure*}

\stitle{Semantic Filters.}
Semantic filter is a core operator in semantic query processing systems. 
Since directly invoking a powerful LLM on every document is often prohibitively expensive, recent studies~\cite{lotus, bargain, taskcascade, scaledoc} have focused on reducing the cost of LLM invocations while preserving high accuracy. These methods follow a two-stage model cascade paradigm: first use a cheaper proxy model to assign each document a proxy score, and then calibrate threshold scores to determine which documents can be confidently labeled by the proxy and which documents should be forwarded to the LLM oracle.
Lotus~\cite{lotus} uses a smaller LLM as the proxy model and learns two thresholds through online sampling. BARGAIN \cite{bargain} optimizes threshold calibration by stratified and progressive sampling for oracle judgment until the target accuracy requirement is satisfied. 
TaskCascade \cite{taskcascade} introduces task decomposition and combines document chunking, task cascade, and model cascade into a unified framework.
ScaleDoc \cite{scaledoc} distills an MLP as the proxy of oracle LLM by contrastive learning, and reconstructs positive and negative distributions from a calibration set for threshold selection.
Distinguished from these methods, our \OASIS adaptively allocates the budgets of training samples to be annotated and the calibration set. Combined with active learning and early-stopping, reduces unnecessary LLM invocations while maintaining high accuracy.


\stitle{LLM-based Semantic Query Processing.}
The strong reasoning capabilities of LLM have enabled growing studies on semantic query processing over unstructured and structured data~\cite{zendb, quest, modora, straptor, unify, lotus, palimpzest, abacus, nivrana, sema}, which aim to optimize the monetary cost of LLMs and the latency of query processing. 
ZenDB \cite{zendb}, QUEST \cite{quest}, Modora \cite{modora}, ST-Raptor \cite{straptor}, and Unify \cite{unify} use LLMs to process unstructured data and exploit data characteristics to reduce the overhead of LLM invocations. 
Lotus \cite{lotus} and Palimpzest \cite{palimpzest} introduce declarative LLM-based semantic operators, analogous to relational operators, but they operate over unstructured data using natural language predicates. 
Abacus \cite{abacus} builds on Palimpzest and proposes cost estimation and Pareto optimization for semantic query planning. 
SEMA ~\cite{sema} embeds semantic operators as first-class citizens into DuckDB, with semantic-expression compression, operator fusion, prompt batching, and adaptive query execution to trade off token cost, latency, and result quality.
Nirvana \cite{nivrana} targets multimodal analytics and optimizes both logical and physical plans to improve accuracy and reduce cost. These studies primarily focus on system-level plan optimization.
In contrast, our \OASIS focuses on online optimization of a fundamental semantic operator, which can be deployed by continuously developing systems. 

\section{Overview of \OASIS}
\label{overview}
We introduce the query specification of semantic filter and the model cascade paradigm in ~\cref{sec:querysepecificationandparadigm}, and the overview of \OASIS in ~\cref{sec:oasisoverview}.

\subsection{Query Specification and Paradigm}
\label{sec:querysepecificationandparadigm}

\OASIS targets processing semantic filter predicates over a large collection of unstructured documents. Let $\mathcal{D}_=\{d_1,\ldots,d_N\}$ denote a document collection, where each document $d_i$ is a text object such as an abstract, report, patent description, or other unstructured record. Unlike conventional value-based predicates, a semantic filter predicate is specified in natural language and requires a semantic understanding of the document content.

\stitle{Semantic Filter.} A semantic filter query is defined as $q=(p,\alpha)$, where $p$ is an online natural-language predicate and $\alpha \in [0,1]$ is a user-defined accuracy target. There is a large language model (LLM) $M$ serving as the semantic oracle, i.e., for a document $d_i \in \mathcal{D}$, the oracle produces a binary judgment: $M(p,d_i) \in \{0,1\}$, where $M(p,d_i)=1$ indicates that $d_i$ satisfies $p$, and $M(p,d_i)=0$ otherwise.
Our task is to identify the documents in $\mathcal{D}$ that satisfy the predicate $p$ and meet the target accuracy.

\stitle{Model Cascade Paradigm.} Directly invoking $M$ on every document produces  oracle-equivalent results, but requires $|\mathcal{D}|$ expensive LLM calls and often incurs unacceptable online latency. This approach cannot scale to large document collections or high-throughput ad hoc workloads. Therefore, existing methods \cite{lotus, bargain, taskcascade, scaledoc} commonly adopt a model cascade paradigm: In the first stage,  a cheaper proxy model assigns each document a proxy score, where a higher proxy score indicates that the document is more likely to satisfy the semantic predicate. In the second stage, an interval $[l,r]$ is selected by a calibration set, such that documents with scores smaller than $l$ or  larger than $r$ are assigned as negative or positive, respectively, while the documents whose scores fall in the uncertainty interval $[l,r]$ are forwarded to the LLM oracle for final judgment.  The key challenge is to deploy an effective proxy model and threshold interval so that the final results satisfy the target accuracy 
$\alpha$ while minimizing expensive oracle invocations.



\subsection{Overview of Three-stage Workflow}
\label{sec:oasisoverview}
Under the model cascade paradigm, \OASIS employs a GNN-based Proxy Model, combining semantic similarities among documents as a query-agnostic prior with the distillation of LLM judgment on a given query. 
Fig.~\ref{fig:overview} presents an overview of \OASIS, which consists of one offline stage and two online stages. 
In the offline stage, \OASIS preprocesses the document collection by extracting the embeddings of the documents via a pre-trained LLM encoder. 
Subsequently, we construct a query-agnostic Navigating Spread-out Graph (NSG) where each node corresponds to a document and edges connect documents with embedding similarities. \OASIS collects a set of initial training samples in the NSG, which will be annotated by any online semantic filter query and used to train the proxy model. The offline preprocessing stage is elaborated on \cref{sec:offline}.

The two online stages of \OASIS follow the general model cascade paradigm while specifically internalizes adaptivity for a given query. 
Specifically, in the first online stage, \OASIS trains a Query-conditioned Graph Attention Network over the document NSG to predict the proxy score of documents, by formulating the task as a semi-supervised node classification. 
\OASIS draws nodes (documents) from the NSG, annotating these samples using the oracle LLM, as the samples to train the model. Furthermore, \OASIS adopts an active learning algorithm that iteratively selects informative nodes with uncertain proxy scores for label annotation, adding these samples to the training set and retraining the model. To avoid unnecessary labeling and training overhead, \OASIS also utilizes an early-stopping mechanism to terminate training once additional labels are unlikely to provide sufficient benefit. We will delve into the model design and training strategy in \cref{sec:model}. 

In the second online stage, after obtaining the proxy scores of all documents, \OASIS calibrates the threshold interval $[l, r]$ with an adaptive size of calibration set. 
Instead of relying on a fixed-size calibration set, \OASIS uses a progressive sampling strategy to estimate the score distribution and determine $[l, r]$. Documents outside the interval can be confidently labeled by the proxy model, whereas documents inside the interval are resort to the oracle LLM. This adaptive threshold calibration strategy further reduces calibration cost while maintaining the target accuracy expectation. We will introduce the details in \cref{sec:calibration}.


\section{Offline Preprocessing}
\label{sec:offline}

As offline preprocessing, \OASIS represents the document-wise similarities as an NSG graph (\cref{sec:offline:graph}), and then identifies anchor nodes (documents) in the graph, which will be annotated using online queries (\cref{sec:offline:annotation}). 

\subsection{Document NSG Construction}
\label{sec:offline:graph}
In information retrieval, k-nearest neighbor (KNN) graph~\cite{dong2011efficient,anastasiu2015l2knng, hnsw, gem} naturally represents pair-wise similarity as a graph, where each node is a document connecting the top-$k$ documents with the highest similarity. However, training a GNN proxy over a naive KNN graph will face some problems. 
For one thing, using a larger $k$ will preserve sufficient but likely redundant  information, resulting in an overly dense graph. Training and evaluating a GNN over a dense graph is not only inefficient for online queries, but also faces the risk of over-smoothing. 
For the other thing,  using a smaller $k$ will miss key connections and degrade the performance of the proxy model. 
In a nutshell, an ideal graph should encode key and sufficient semantic similarities among documents and maintain a sparse topology for efficient deployment.

Based on these considerations, \OASIS constructs a Navigating Spread-out Graph (NSG)~\cite{nsg, kim2026worksharingoffloadingefficient} over a collection of documents. Specifically, \OASIS uses a state-of-the-art pre-trained embedding model to transform each document $d_i \in \mathcal{D}$ into a $n_d$-dimensional embedding $\mathbf{e}_i \in \mathbb{R}^{n_d}$. The similarity between two documents is measured by the cosine similarity of their embeddings, and each document, serving as a node of graph, possesses a candidate neighbor set containing its top-$k$ similar nodes. 
For each node $d_i$, \OASIS collects its 1-hop and 2-hop candidate neighbors, reranking these neighbors by their similarities to $d_i$ and applying NSG pruning~\cite{nsg} on these neighbors: a candidate neighbor of node $d_i$ is pruned if it is more similar to an existing neighbor than to $d_i$. 
To this end, redundant edges are removed from the graph while the connectivity and navigability of semantic similarity are preserved, allowing efficacy of GNN message passing and the efficiency of model training.

\subsection{Sampling Anchor Nodes for Annotation}
\label{sec:offline:annotation}

After constructing the document NSG, we need to sample a small number of documents for LLM oracle annotation. Once a query $q = (p, \alpha)$ arrives, each document $d_i$ is labeled online by the oracle LLM as $L_i = M(p, d_i) \in \{0, 1\}$. These labeled documents serve as initial anchor nodes and provide semi-supervised signals for training the GNN.  Instead of sampling documents uniformly, which may over-sample dense local regions, \OASIS collects documents from different regions of the document NSG, balancing coverage and diversity~\cite{coreset1, coreset2}. Given the document NSG $G =(V, E)$, where $V$ is the node set and $E$ is the edge set, a budget $B$, \OASIS identifies $B$ distinct anchor nodes as a set $\mathcal{O}$ around $m$ centroid nodes. 

First, \OASIS selects these centroids $C=\{c_1,\ldots,c_m\}$, using farthest-point sampling over graph distance. The first center is selected randomly. Then, for $i = 2, \cdots, m$, the next centroid $c_i$ is selected as one of the nodes farthest from the existing centroids $C_{i-1} = \{c_1, \cdots, c_{i-1} \}$:
\begin{equation}
    c_{i} \in \arg\max_{v\in V \setminus C_{i-1}} \min_{c\in C_{i-1}} \text{dist}_G(v,c),
\end{equation}
where $\text{dist}_G$ denotes the shortest distance between two nodes in $G$.  This greedy procedure tends to select centroids spanning different regions of the document graph.
After $C$ is determined, each node $v$ is assigned to a cluster $R_j$ of its nearest centroids, i.e., $R_j = \{v\in V \mid c_j = \arg\min_{c\in C} dist_G(v,c)\}$.
The resulting clusters can be viewed as graph-based Voronoi cells, where each cluster contains documents that are closest to the same centroid. 
Finally, we sample nodes from $\{R_1, \cdots, R_m\}$ to the anchor node set $\mathcal{O}$ without replacement.

\stitle{Computational Complexity.}
The time complexity of constructing the document NSG is $O(N^2n_d + Nk^2rn_d)$, where $n_d$ is the embedding dimension, and $r$ is the maximum number of neighbors for each node after pruning. The time complexity of anchor node sampling is  $O(m(N+|E|)+B)$. 

\section{GNN-Based Proxy Model}
\label{sec:model}

\OASIS employs a GNN-based Proxy Model, combining semantic similarities among documents as a query-agnostic prior with the distillation of LLM judgment on a given query.
\cref{sec:model:architecture} presents the model architecture and training objectives. The model is further optimized by online incremental sampling and active learning, as shown in \cref{sec:model:active}.

\subsection{Query-conditioned GAT as a Proxy}
\label{sec:model:architecture}

Given an online query $q$, we formulate the construction of the proxy model as a semi-supervised node classification task in the NSG graph, where the ground-truth labels are obtained by annotating the anchor  nodes $\mathcal{O}$ via online LLM oracle invocations. 
\OASIS employs a Graph Attention Network (GAT)~\cite{velickovic2018graph, transformer} as a query-specific proxy model, and the model is trained to predict a proxy score $\hat{y}_i$ for the remaining nodes $v_i \in V \setminus \mathcal{O}$. 
The graph structure of the underlying NSG provides a query-agnostic prior of semantic correlation between documents. In the graph, learning-based message passing propagates the oracle signals of anchor nodes to the remaining nodes, which are learned to adapt to a specific query. 

\subsubsection{Query-condition GAT}

Given the document NSG, each node $v_i$ is set to an initial embedding $\mathbf{h}_i^{(0)}$, combining its possible oracle label, the document embedding $\mathbf{e}_i$, and the similarity between document and the query as:
\begin{equation}
\mathbf{h}_i^{(0)}
=
s_i \cdot \mathbf{W} \cdot \mathbf{{e}}_i+ \mathbf{emb}(a_i)
,~~
a_i = \begin{cases}
    -1~\text{or}~1 & \text{if}~v_i \in \mathcal{O}, \\
    0 & \text{otherwise}, 
\end{cases}
\end{equation}
where $s_i$ is the cosine similarity between $\mathbf{e}_i$ and the embedding of the query, $\mathbf{W}$ is a weight matrix. The query similarity acts as a soft gate over the document representation~\cite{querygated, transformer, bert}, allowing documents that are more relevant to the current predicate query to contribute stronger semantic signals.
The scalar $a_i \in \{-1, 0, 1\}$ denotes the possible oracle label and is 0 if the node is not an anchor, which is transformed into a vector by a learnable embedding function $\mathbf{emb}(\cdot)$.
The model employs $L_{num}$ GAT layers, where a node aggregates messages from its neighbors through attention-weighted summation with a residual connection to the initial embedding $\mathbf{h}_i^{(0)}$.
To capture information from different propagation depths, the model fuses the final node embedding using three multi-depth branches over the initial, middle-layer, and final-layer embeddings \cite{tempoalbranches}:
\begin{equation}
\mathbf{z}_i = 
\phi_e\left(\mathbf{h}_i^{(0)}\right)
\Vert
\phi_m\left(\mathbf{h}_i^{(\lfloor L_{num}/2 \rfloor)}\right)
\Vert
\phi_l\left(\mathbf{h}_i^{(L_{num})}\right), 
\end{equation}
where $\phi_e$, $\phi_m$, and $\phi_l$ are linear transformations, and $\Vert$ denotes vector concatenation. This fused embedding $\mathbf{z}_i$ is transformed into the proxy score $\hat{y}_i \in [0, 1]$ by a multilayer perceptron (MLP) as
\begin{equation}
\label{eq:node_represention}
\hat{y}_i = \text{Sigmoid}(o_i), \quad o_i =\mathrm{MLP}(\mathbf{z}_i). 
\end{equation}

\begin{figure}[t]
    \centering
    \includegraphics[scale=0.41]{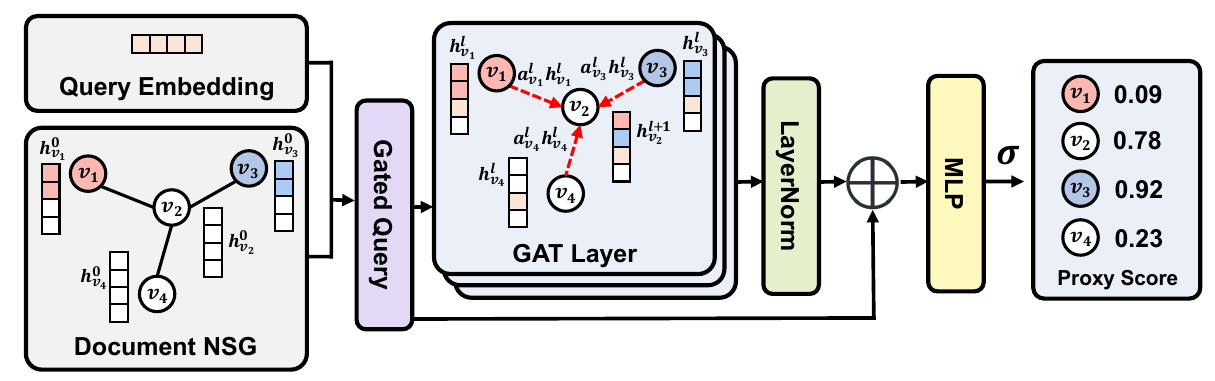}
    
    \caption{Illustration of query-conditioned graph attention network (The branches are omitted for simplicity).}
    \label{fig:gnn}
\end{figure}

\subsubsection{Training Objectives}
The training objective of the proxy is designed to produce a well-separated score distribution: positive documents obtain scores close to 1, while negative documents obtain scores close to 0, yielding an approximately bimodal distribution. 
Given the anchor nodes set $\mathcal{O}$, each anchor node $v_i \in \mathcal{O}$ obtains a ground-truth label $y_i \in \{0, 1\}$ via oracle annotation.
\OASIS optimizes the GAT proxy using three loss terms as follows.

\stitle{Binary cross-entropy Loss.}
For labeled anchors, we first apply binary cross-entropy (BCE) to fit the oracle labels, encouraging positive anchors to have scores close to 1 and negative anchors to have scores close to 0. To handle class imbalance in the sampled anchor set, we use a positive-class weight as $w^+={|\mathcal{O}^-|} / {|\mathcal{O}^+|}$, where $|\mathcal{O}^+|$ and $|\mathcal{O}^-|$ are the numbers of positive and negative anchors, respectively. The weighted BCE loss for anchor classification is:
\begin{equation}
\mathcal{L}_{\mathrm{bce}}
=
-
\frac{1}{|\mathcal{O}|}
\sum_{i\in\mathcal{O}}
\left[
w^+ y_i \log \hat{y_i}
+
(1-y_i)\log(1- \hat{y_i})
\right].
\end{equation}
To prevent model overconfidence on limited anchor supervision, 
we apply label smoothing \cite{loss1} on $y_i$, i.e., ${y}_i=(1-\epsilon)y_i+0.5\epsilon$, with a smoothing coefficient $\epsilon \in (0, 1)$.

\stitle{Supervised  Contrastive Loss.}
To improve the separability of positive and negative anchor nodes in the latent  space, we additionally use a supervised contrastive loss over the fused anchor embeddings~\cite{loss2}. Let $\bar{\mathbf{z}}_i={\mathbf{z}_i} / {\|\mathbf{z}_i\|_2}$ be the normalized representation of $\mathbf{z}_i$ (in Eq.~\eqref{eq:node_represention}) for an anchor $v_i \in \mathcal{O}$. The pairwise similarity between two anchors is computed as $s_{ij}={\bar{\mathbf{z}}_i^\top \bar{\mathbf{z}}_j}/{\tau}$
where $\tau$ is the temperature. For each anchor node $v_i$, we define its positive set as all the remaining nodes with the same label, i.e., $\mathcal{P}(i)=\{v_j\in\mathcal{O}\mid j\ne i, y_j=y_i\}$. The supervised contrastive loss is
\begin{equation}
    \mathcal{L}_{\mathrm{sup}}
=
-
\frac{1}{|\mathcal{O}|}
\sum_{v_i\in\mathcal{O}}
\frac{1}{|\mathcal{P}(i)|}
\sum_{v_j\in\mathcal{P}(i)}
\log
\frac{
\exp(s_{ij})
}{
\sum_{v_k\in\mathcal{O}, k\ne i}
\exp(s_{ik})
}
\end{equation}
This contrastive loss pulls same-label anchors closer in the embedding space while pushing different-label anchors apart, thereby improving the separability of positive and negative documents.

\stitle{Graph Smoothness Loss.}
 Finally, we impose a graph smoothness regularizer on adjacent non-anchor nodes to prevent nodes within the same semantic region from exhibiting random fluctuations~\cite{loss3}. This regularizer discourages sharp local spikes in the score distribution and promotes locally consistent predictions.
Let $E'=\{(v_i, v_j) \in E \mid v_i, v_j \in V \setminus \mathcal{O}\}$ denote the set of all pairs of adjacent non-anchor nodes.

 The smoothness loss penalizes large differences in logits between adjacent nodes in $E'$:
\begin{equation}
\mathcal{L}_{\mathrm{reg}}
=
\frac{1}{|E'|}
\sum_{(v_i, v_j)\in E'}
(o_i-o_j)^2    
\end{equation}

The final objective is a weighted combination of the three loss terms:
\begin{equation}
\mathcal{L}
=
\mathcal{L}_{\mathrm{bce}}
+
\lambda_{\mathrm{sup}}\mathcal{L}_{\mathrm{sup}}
+
\lambda_{\mathrm{reg}}\mathcal{L}_{\mathrm{reg}},
\end{equation}
where $\lambda_{\mathrm{sup}}$ and $\lambda_{\mathrm{reg}}$ are hyperparameters that control the weights of the contrastive term and the smoothness regularizer, respectively. Overall, this objective enables the proxy model to fit sparse anchor labels, separate positive and negative evidence in the embedding space, and produce locally consistent predictions over unlabeled neighborhoods.

\subsection{Active Learning}
\label{sec:model:active}

The quality of the GAT proxy depends not only on the model architecture, but also on which documents are annotated as anchors. A fixed one-shot offline  sampling strategy only provides limited anchor nodes with diverse document features, which may be insufficient to provide supervision for hard online queries. To address this issue, \OASIS adopts an active learning strategy that incrementally selects informative documents for LLM annotation. 
At each iteration, \OASIS trains the proxy model with the currently labeled anchors, inspects the resulting score distribution, and samples additional documents from regions that are likely to improve the decision boundary.  To further control the online budget, \OASIS incorporates early-stopping criteria that terminate the active learning loop once additional labeling is unlikely to yield sufficient benefit. We first present the overall workflow, followed by the incremental sample-selection strategy and the early-stopping mechanisms.


\begin{algorithm}[!t]
\footnotesize
\caption{\textsc{ActiveLearningFramework}}
\label{alg:overview}
\KwIn{$\mathcal{D}$: document collection, $B_{ini}$: initial budget,
$\Delta B$: incremental budget, $B_{total}$: total budget,
$M$: Oracle LLM, $q$: online filter query}
\KwOut{$P$: \add{predicted score distribution}}

$\mathcal{G} \gets$ \textsc{ConstructDocumentNSG}($\mathcal{D}$)\tcp*[r]{Performed offline}
$B \gets \min(B_{ini}, B_{total})$\;
$\mathcal{O} \gets$ \textsc{InitialSampleAndAnnotate}($\mathcal{G}, B, q, M$)\;

\While{\textbf{true}}{
    $P \gets$ \textsc{GNNTrainingAndInference}($\mathcal{G}, q, \mathcal{O}$)\;
    $I \gets$ \textsc{IdentifyValleyInterval}($P$)\;
    \If{$B = B_{total}$ \textbf{or} $\mathrm{Stop}_{II}(P)$
        \textbf{or} $\mathrm{Stop}_{I}(P,I)$}{
        break\;
    }
    \If{$I = \varnothing$}{
        $I \gets \mathrm{range}(P)$\;
    }
    $\delta \gets \min(\Delta B, B_{total}-B)$\;
    $\Delta\mathcal{O} \gets$
    \textsc{StratifiedSampleAndAnnotate}($P,I,\delta,M$)\;
    $\mathcal{O} \gets \mathcal{O}\cup\Delta\mathcal{O},
    \quad B \gets B+\delta$\;
}
\Return{$P$}\;
\end{algorithm}

\subsubsection{Algorithm Overview}
The main idea of the active learning algorithm is iterative training with incremental sampling. Instead of labeling a fixed number of documents at once, \OASIS starts from the initial anchor set and then adds a small batch of newly selected documents in each round. These new documents are selected according to the score distribution produced by the proxy model from the previous round.
Alg. ~\ref{alg:overview} presents the workflow for processing an online semantic filter query in proxy scoring stage. After the document NSG has been constructed offline and the initial anchors have been sampled and annotated by the LLM oracle (lines 1-2), the algorithm proceeds iteratively. In each iteration, \OASIS first trains the query-conditioned GAT using the currently accumulated anchor set and performs inference over all documents (line 5). This produces a proxy score distribution over the document collection (line 6).
Based on this distribution, \OASIS then identifies a potential valley interval, which corresponds to the low-density region between two score modes (line 9). Documents in this interval are likely to be close to the current decision boundary and are therefore informative for refining the proxy model. If a reliable valley interval is detected, \OASIS performs stratified sampling within this interval and sends the selected documents to the LLM oracle for annotation (line 14). Otherwise, when the score distribution has not yet formed a clear bimodal structure, \OASIS falls back to stratified sampling over the full score range to preserve exploration (lines 12-13).
The loop terminates when either the total sampling budget $B_\text{total}$ exhausted, the ambiguous region becomes sufficiently small (lines 10-11), or the score distribution converges across iterations (lines 7-8). In this way, \OASIS adaptively allocates the LLM labeling budget according to the difficulty of the current query.

\subsubsection{Incremental Sampled Data Selection}
\label{dataselection}

Active learning requires the system to identify informative unlabeled samples and acquire oracle labels for subsequent model refinement~\cite{coreset1, activelearning1}.  In our semantic filtering task, the current GNN proxy assigns each document a score indicating how likely the document is to satisfy the predicate. Ideally, after sufficient training, the score distribution becomes approximately bimodal: negative documents concentrate low scores close to 0, while positive documents concentrate high scores close to 1. 
Under such a bimodal distribution, the most informative documents are those located in the ambiguous region between the two modes. These documents are close to the proxy decision boundary and are therefore more useful for improving the model than high-confidence samples. 

However, identifying this ambiguous region is non-trivial in the online scenario. First, ground-truth labels are unavailable unless the LLM oracle is invoked, so label-based uncertainty metrics ~\cite{2017nips} cannot be directly used. Second, sampling around a fixed threshold, e.g., 0.5, is unreliable, because proxy scores may be biased or poorly calibrated, especially in early iterations. Therefore, \OASIS identifies the ambiguous region solely from the landscape of the unlabeled score distribution, without requiring calibrated probabilities or additional oracle labels.

\stitle{Density-aware Uncertain Interval Detection.}
At active learning iteration $t$, \OASIS first estimates the density of the proxy scores and applies smoothing to suppress local noise. Let $\tilde{f}^{(t)}(\cdot)$ denote the smoothed score density function at iteration $t$. \OASIS searches for a peak-valley-peak pattern in this density curve. 
For two adjacent significant peaks $p_l$ and $p_r$ detected by SciPy library~\cite{scipy2020}, the valley position between them is defined as:
\begin{equation}
v_c=\arg\min_{p_l<b<p_r}\tilde{f}^{(t)}(b).
\end{equation}
Here, $p_l$ and $p_r$ denote the left and right score modes, and $v_c$ denotes the lowest-density position between them. 
Intuitively, documents whose scores fall around $v_c$ lie between the positive and negative groups and are therefore likely to be informative for refining the decision boundary.

\begin{figure}[t]
    \centering
    \includegraphics[scale=0.5]{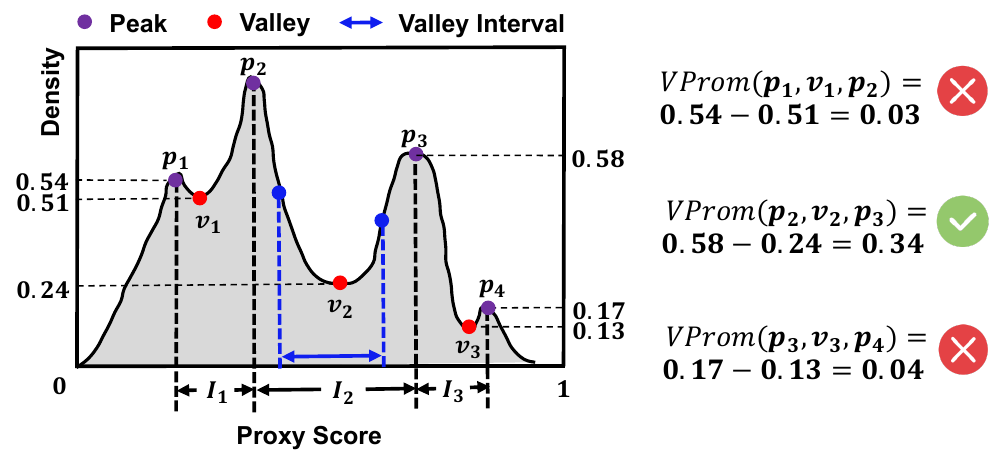}
   
    \caption{Example of uncertain interval detection.}
    \label{fig:valley}
\end{figure}

Inspired by peak prominence in signal processing \cite{scipy2020} and density-valley separation used in clustering \cite{2014science, dbscan, 2013dbscan}, \OASIS evaluates the reliability of a candidate peak-valley-peak structure by the valley prominence with respect to its weaker neighboring mode:

\begin{equation} \label{eq1}
    \mathrm{VProm}(p_l,v_c,p_r) = \min\left(\tilde{f}^{(t)}({p_l}),\tilde{f}^{(t)}({p_r}) \right) - \tilde{f}^{(t)}({v_c}).
\end{equation}
This score measures the vertical separation between the valley and the lower of its two adjacent peaks. Using the lower peak gives a conservative estimate of bimodal separation and avoids selecting shallow valleys adjacent to insignificant modes.

In addition, \OASIS requires the relative valley depth to satisfy
\begin{align}
    \frac{\tilde{f}^{(t)}({v_c})}{\min \left(\tilde{f}^{(t)}({p_l}),\tilde{f}^{(t)}({p_r})\right)} < \tau_{val}, 
\end{align}
where $\tau$ is a predefined threshold of relative depth. This condition ensures that the selected valley is sufficiently deep relative to its adjacent modes. Among all valid candidate valleys, \OASIS selects the one with the largest valley prominence.

After selecting the valley position $v_c$, \OASIS expands the valley into an  interval as the uncertain region of proxy score. We define the density threshold $\theta^{(t)}$ for interval expansion as  
\begin{align}\label{eq2}
\add{\theta^{(t)} = 
\tilde{f}^{(t)}({v_c})
+
\alpha_{exp}\left(\min \left(\tilde{f}^{(t)}({p_l}),\tilde{f}^{(t)}({p_r}) \right)-\tilde{f}^{(t)}({v_c})\right)}, 
\end{align}
where $\alpha_{exp} \in [0, 1]$ controls how far the interval expands from the valley toward the adjacent peaks. 
Starting from $v_c$, \OASIS expands left and right until the smoothed density exceeds $\theta^{(t)}$. The resulting uncertain interval $I^{(t)} = [l^{(t)}, r^{(t)}]$ covers the low-density region around the valley while excluding the high-confidence peak regions.

\begin{example}
    Fig. \ref{fig:valley} shows an example of selecting valley interval. We first identify 4 peaks and the 3 valley points between adjacent peaks, obtaining three candidate intervals $I_1$, $I_2$, and $I_3$. We then compute the score of each interval according to Eq.~\eqref{eq1}. The score of $I_2$ is 0.34, which is the largest among the three intervals. Therefore, $I_2$ is more likely to be the ambiguous region between positive and negative instances in the mixed distribution, while $I_1$ and $I_3$ are more likely to be spikes. Finally, we expand from $v_2$ to both sides and obtain the valley interval according to Eq.~\eqref{eq2}, which is shown by the blue arrows.
\end{example}

\stitle{Stratified Sample within Uncertain Interval.} 
After obtaining the uncertain interval $I^{(t)}$, \OASIS samples a new batch of documents from the unlabeled nodes whose proxy scores fall inside this interval.
\OASIS partitions $I^{(t)}$ into score bins and performs stratified sampling across these bins. The number of sampled documents from each bin is proportional to the number of candidate documents in that bin, while the total number of newly selected samples is limited to the incremental budget $\Delta B$. 
If no reliable uncertain interval is detected,  \OASIS falls back to stratified sampling over the full score range. \add{The uncertain interval and full range are partitioned into 10 and 20 equal-width bins, respectively.} This fallback is important in early iterations, when the proxy model may not yet produce a clear bimodal distribution. It preserves exploration and allows the model to collect labels from diverse score regions. All selected documents are then sent to the LLM oracle for annotation, and the returned labels are added to the anchor set for the next training iteration.


\subsubsection{Early-Stopping Strategies}
Although incremental sampling improves proxy quality, continuing active learning  can waste LLM calls once the score distribution has become sufficiently stable. \OASIS therefore uses two early-stopping strategies to avoid unnecessary labeling and retraining.

\stitle{Strategy I: by Uncertain Documents Ratio.}
The first early-stopping criterion is triggered when the number of documents in the ambiguous region becomes sufficiently small  (line 10 in Algorithm \ref{alg:overview}). After identifying the uncertain interval $I^{(t)}$, \OASIS counts the number of documents whose proxy scores fall inside this interval: $N^{(t)}_\text{unc} = |\{ v_i \in V \mid \hat{y}^{(t)}_i \in I^{(t)} \}|$, where $\hat{y}^{(t)}_i$ is the model prediction of document $v_i$ at learning iteration $t$. 
If the ratio of these uncertain documents $N^{(t)}_\text{unc}/|V|$ is smaller than a predefined threshold $\eta_1 \in [0, 1]$, \OASIS stops active learning. 
The intuition is that when only a small fraction of documents remains near the decision boundary, further proxy refinement benefits limited documents and may even risk model overfitting.

\stitle{Strategy II: by Score Distribution.}
The second early-stopping criterion is triggered when the score distribution converges across iterations (line 7 in Algorithm \ref{alg:overview}). \OASIS uses two label-free stability measures.
First, \OASIS measures the average score change between two consecutive iterations:
\begin{align}
    \Delta^{(t)}_\text{score} = \frac{1}{|V|} \sum_{v_i \in V} |\hat{y}^{(t)}_i - \hat{y}^{(t-1)}_i|
\end{align}
A small value of $\Delta^{(t)}_\text{score}$ indicates that the proxy scores have stabilized. Second, \OASIS measures the stability of the score ranking on a set of uniformly sampled document pairs $\mathcal{R}$ by computing the rank-flip ratio:
\begin{align}
    \Delta^{(t)}_\text{rank} \frac{1}{|\mathcal{R}|} \sum_{(v_i, v_j) \in \mathcal{R}} \mathds{1} \left[ \left( \hat{y}^{(t)}_i  - \hat{y}^{(t)}_j \right) \left( \hat{y}^{(t-1)}_i  - \hat{y}^{(t-1)}_j \right) <0 \right].
\end{align}

This quantity measures the fraction of sampled document pairs whose relative ordering changes between two adjacent iterations. If both $\Delta^{(t)}_\text{score} < \eta_2$ \emph{and} $\Delta^{(t)}_\text{rank} < \eta_3$, \OASIS considers the proxy score distribution to have converged and stops active learning. Here, $\eta_2, \eta_3 \in [0, 1]$ are predefined thresholds.

\stitle{Complexity Analysis.}
The time complexity of proxy scoring is $O\left( TQL(Nn_d^2+|E|n_d) + TQ B_{\mathrm{total}}^2n_d + T(N+|\mathcal R|) + \sum_{d_i \in \mathcal{O}} C_M(d_i)\right) $, where $T$ is the iteration round number of active learning, $Q$ is the epoch number of GAT training in each round, and $C_M(d_i)$ is the LLM labeling cost for document $d_i$. 

\section{Adaptive Threshold Calibration}
\label{sec:calibration}

After obtaining the final proxy score for each document, \OASIS selects a threshold interval $[l,r]$, within which documents are treated as ambiguous and processed by the LLM oracle. Existing calibration-based cascade methods typically adopt fixed-size sampling to estimate the score-label distribution. For example, ScaleDoc discretizes the score range into bins and samples a fixed size of documents in each bin as calibration set, and acquires their binary labels by LLM oracle and reconstructs the positive and negative score distributions, and then selects the filtering thresholds from bin boundaries that minimize the unfiltered rate while satisfying the target accuracy constraint~\cite{scaledoc}. However, a fixed calibration ratio is suboptimal for ad hoc semantic predicates, since different queries may induce substantially different score distributions. For easy queries, positive and negative documents are well separated in the proxy-score space, so the distribution reconstruction and threshold selection are reliable with only a small number of labels. In contrast, for hard queries, the two distributions may heavily overlap around the decision boundary, making the selected thresholds highly sensitive to sampling noise.

This observation can be formalized from a sample-complexity perspective. Let the score range be discretized into a finite threshold search space $\Theta$, and let $\theta^*$ denote the optimal threshold pair obtained on the full document set under the accuracy constraint. We have the following proposition.

\begin{proposition}
\label{pro:overlap_complexity}
Let the proxy score range be partitioned into $H$ bins
$\{\mathcal D_h\}_{h=1}^{H}$, where
$N_h=|\mathcal D_h|$, $w_h=N_h/N$, and $y_i \in \{0,1\}$ is the ground-truth label of document $d_i$. Let $\eta_h=\frac{1}{N_h}\sum_{d_i\in\mathcal D_h} y_i$
denote the positive-label proportion in bin $\mathcal D_h$, and define
the score-label overlap complexity as
$\mathcal V_{\mathrm{ov}}=\sum_{h=1}^{H}w_h\eta_h(1-\eta_h).$
Suppose that a calibration set is constructed by uniformly sampling $m_h=\rho N_h$ documents from each bin, and let
$\widehat{Acc}_{\rho}(\theta)$ be the estimated accuracy of a threshold pair $\theta\in\Theta$ based on the sampled labels. Then, for any $\epsilon>0$,
$
\Pr\left(
\sup_{\theta\in\Theta}
\left|
\widehat{Acc}_{\rho}(\theta)-Acc(\theta)
\right|
\ge \epsilon
\right)
\le
2|\Theta|
\exp\left(
-\frac{\rho N\epsilon^2}
{2(\mathcal V_{\mathrm{ov}}+\epsilon/3)}
\right).
$
Consequently, to estimate the accuracies of all candidate threshold
pairs within $\epsilon$ with probability at least $1-\delta$, it is
sufficient to use 
$
\rho
\ge
\frac{
2(\mathcal V_{\mathrm{ov}}+\epsilon/3)
}{
\epsilon^2N
}
\log\frac{2|\Theta|}{\delta}.
$
\end{proposition}

\begin{proof}
For a threshold pair $\theta=[l,r]$ where $l$ and $r$ are bin boundaries, let $\mathcal H^-_\theta$, $\mathcal H^0_\theta$, and $\mathcal H^+_\theta$ denote the sets of score bins below $l$, within $[l,r]$, and above $r$, respectively. Documents in $\mathcal H^-_\theta$ are assigned negative labels, documents in $\mathcal H^+_\theta$ are assigned positive labels, and documents in $\mathcal H^0_\theta$ are processed by the LLM oracle. Therefore, $Acc(\theta)
=
\sum_{h\in\mathcal H^-_\theta}
w_h(1-\eta_h)
+
\sum_{h\in\mathcal H^0_\theta}
w_h
+
\sum_{h\in\mathcal H^+_\theta}
w_h\eta_h.$ Let $\widehat{\eta}_h$ be the positive-label proportion estimated from the $m_h=\rho N_h$ calibration samples in bin $\mathcal D_h$. The estimated accuracy is $\widehat{Acc}_{\rho}(\theta)
=
\sum_{h\in\mathcal H^-_\theta}
w_h(1-\widehat{\eta}_h)
+
\sum_{h\in\mathcal H^0_\theta}
w_h
+
\sum_{h\in\mathcal H^+_\theta}
w_h\widehat{\eta}_h.$ Define $c_h(\theta)=-1$ if $h\in\mathcal H^-_\theta$, $c_h(\theta)=0$ if $h\in\mathcal H^0_\theta$, and $c_h(\theta)=1$ if $h\in\mathcal H^+_\theta$. Then, $\widehat{Acc}_{\rho}(\theta)-Acc(\theta)=\sum_{h=1}^{H}w_hc_h(\theta)\bigl(\widehat{\eta}_h-\eta_h\bigr)$.
Since $m_h=\rho N_h$, we have $\frac{w_h}{m_h}=\frac{N_h/N}{\rho N_h}=\frac{1}{\rho N}$. Expanding each stratified estimator into its sampled observations, the estimation error can be written as a sum of centered bounded random variables. Specifically, each sample-level random variable is bounded in absolute value by $1/(\rho N)$. Its total variance is bounded by $\sum_{h=1}^{H}\frac{w_h^2}{m_h}\eta_h(1-\eta_h)=\frac{1}{\rho N}\sum_{h=1}^{H}w_h\eta_h(1-\eta_h)=\frac{\mathcal V_{\mathrm{ov}}}{\rho N}$.
Applying Bernstein's inequality for stratified sampling, for any fixed $\theta\in\Theta$, gives $\Pr\left(\left|\widehat{Acc}_{\rho}(\theta)-Acc(\theta)\right|\ge\epsilon\right)\le 2\exp\left(-\frac{\rho N\epsilon^2}{2(\mathcal V_{ov}+\epsilon/3)}\right)$.
Applying the union bound over all $\theta\in\Theta$ yields \\ $\Pr\left(\sup_{\theta\in\Theta}\left|\widehat{Acc}_{\rho}(\theta)-Acc(\theta)\right|\ge\epsilon\right)\le 2|\Theta|\exp\left(-\frac{\rho N\epsilon^2}{2(\mathcal V_{ov}+\epsilon/3)}\right)$. Setting the right-hand side to be at most $\delta$ gives the stated sample-size condition.
\end{proof}

Pro. \ref{pro:overlap_complexity} reveals an important query-adaptive property: the required calibration sample size depends on the overlap complexity $\mathcal{V}_{ov}$.
Easy queries with lower $\mathcal{V}_{ov}$ lead to less calibration samples. However, $\mathcal{V}_{ov}$ cannot be known in advance because the full ground-truth labels are unavailable in the online phase. To address this challenge, we propose a progressive adaptive calibration sampling strategy which is based on the following proposition.

\begin{proposition}
\label{pro:threshold_stability}
\add{Let $\alpha$ be the target accuracy, and let
$\theta^*$ and $\widehat{\theta}_m$ denote the
optimal threshold pairs selected using the full
data and $m=\rho N$ calibration samples,
respectively. Define the query-specific
accuracy margin as
$\gamma=\min_{\theta\in\Theta}|Acc(\theta)-\alpha|>0,$ we have
$\Pr(\widehat{\theta}_m\neq\theta^*)\le2|\Theta|\exp\left(-\frac{m\gamma^2}{2(\mathcal V_{\mathrm{ov}}+\gamma/3)}\right).$}
\end{proposition}

\begin{proof}
When the uniform estimation error is smaller
than $\gamma$, all candidate thresholds retain
their true feasibility status. Since the
unfiltered rate is independent of calibration
labels, the feasible set and optimization
objective remain unchanged, yielding
$\widehat{\theta}_m=\theta^*$ under the same
tie-breaking rule. The result follows from
Pro. \ref{pro:overlap_complexity} with
$\epsilon=\gamma$.
\end{proof}

\add{Pro. \ref{pro:threshold_stability} shows that increasing calibration samples improves the reliability of threshold selection, making the selected thresholds increasingly stable. Motivated by this observation, \OASIS progressively increases calibration samples and monitors threshold convergence as a practical stopping signal.}

Specifically, we first randomly sample $a_{\mathrm{ini}}\%$ of documents
within each bin, and then run the distribution reconstruction and threshold
selection algorithm in~\cite{scaledoc} to obtain the initial threshold
interval $I_{cas}^{(0)}=[l_0,r_0]$. At the $t$-th iteration, we incrementally sample an
additional $\Delta a\%$ fraction of documents from each bin, combine them
with the previously accumulated calibration samples, reconstruct the
positive and negative score distributions again, and compute a new threshold
interval $I_{cas}^{(t)}=[l_t,r_t]$. \add{Notably, the samples used for calibration are disjoint from those used for proxy training.} We repeat this process until the selected thresholds
converge. In this way, the calibration budget is automatically
adapted to the difficulty of each query: easy queries can stop with fewer
oracle labels, while hard queries receive more calibration samples to obtain
stable thresholds. The following proposition explains why threshold stability provides a meaningful stopping signal.

\begin{proposition}
\label{pro3}
Let $A\subset B\subseteq\mathcal{D}$ be two nested calibration
sets, and suppose that the same threshold pair $\theta=[l,r]$ is
acquired under both sets. Let
$\mathcal{D}_\theta=\{d_i:l\le \hat{y}_i\le r\}$ be the ambiguous documents set, and let
$E_\theta=
\{d_i\in\mathcal{D}\setminus \mathcal{D}_\theta:
Predict(\hat y_i)\neq y_i\}$ be the documents outside the ambiguous region that will be misjudged by the proxy. Then the LLM reduction rate and accuracy satisfy
$
R_A-R_B
=
\frac{
|(B\setminus A)\cap(\mathcal{D}\setminus \mathcal{D}_\theta)|
}{N}
$
and
$
Acc_B-Acc_A
=
\frac{|(B\setminus A)\cap E_\theta|}{N}.
$
Consequently,
$
0\le Acc_B-Acc_A\le R_A-R_B.
$
\end{proposition}
 
\begin{proof}
 Since the same threshold pair $\theta=[l,r]$ is selected under $A$ and
$B$, the ambiguous document set $D_\theta$ is identical in the two
cases. Under calibration set $A$, the LLM is invoked for every document in
$A$ during calibration and every ambiguous document that has not already
been labeled. Since calibration labels are reused, the total number of
distinct LLM invocations is $C_A
=
|A|+|\mathcal{D}_\theta\setminus A|.$ Similarly, under calibration set $B$, $C_B
=
|B|+|\mathcal{D}_\theta\setminus B|.$ Because $A\subset B$, we have
$
\mathcal{D}_\theta\setminus A
=
(\mathcal{D}_\theta\setminus B)
\mathbin{\dot\cup}
\bigl(\mathcal{D}_\theta\cap(B\setminus A)\bigr),
$ where $\dot\cup$ denotes a disjoint union. Therefore,
$C_B-C_A=|B\setminus A|-|\mathcal{D}_\theta\cap(B\setminus A)|=
|(B\setminus A)\cap(\mathcal{D}\setminus \mathcal{D}_\theta)|.$ Dividing by $N$ gives
$R_A-R_B=
\frac{
|(B\setminus A)\cap
(\mathcal{D}\setminus \mathcal{D}_\theta)|
}{N}.$ Next, all documents in $\mathcal{D}_\theta$ are processed by the LLM oracle and are
therefore correctly classified. Documents in the calibration set are also
labeled by the LLM oracle and correctly classified. Hence, under calibration set
$A$, the only remaining errors are the documents outside the
ambiguous region that will be misclassified by the proxy model and are not contained in $A$. Thus,
$Acc_A=1-\frac{|E_\theta\setminus A|}{N}.$ Similarly,
$Acc_B=1-\frac{|E_\theta\setminus B|}{N}.$ Since $A\subset B$,
$E_\theta\setminus A
=
(E_\theta\setminus B)
\mathbin{\dot\cup}
\bigl(E_\theta\cap(B\setminus A)\bigr).$ It follows that
$Acc_B-Acc_A=\frac{|(B\setminus A)\cap E_\theta|}{N}.$ Finally, by definition,
$E_\theta\subseteq\mathcal{D}\setminus \mathcal{D}_\theta.$ Therefore,
$
|(B\setminus A)\cap E_\theta|
\le
|(B\setminus A)\cap
(\mathcal{D}\setminus \mathcal{D}_\theta)|.$ Combining the two results yields $0\le Acc_B-Acc_A\le R_A-R_B.$
\end{proof}

Pro. \ref{pro3} shows that if the thresholds selected from a smaller set A and a larger set B are the same, they only increase the oracle cost for samples outside the ambiguous interval, while improving accuracy only for the subset that would otherwise be misjudged by the proxy model. Therefore, the marginal accuracy gain is upper-bounded by the normalized marginal oracle cost, indicating that the smaller calibration set is more cost-effective.

\begin{table}[t]
\footnotesize
\centering
\caption{Statistics and example queries of the datasets.}
\label{tab:dataset_statistics}
\begin{tabularx}{\columnwidth}{l c X}
\toprule
\textbf{Dataset} & \textbf{Avg. \# Words} & \textbf{Example Query} \\
\midrule
\textsc{PubMed}
& 328 
& Does the study indicate a positive effect of the treatment? \\

\textsc{BigPatent}
& 118 
& Does the invention combine sensing, monitoring, or feedback with an active response or adjustment? \\

\textsc{GovReport} 
& 498 
& Does the report find that an agency made progress but still has unresolved implementation challenges? \\
\bottomrule
\end{tabularx}
\end{table}

\begin{figure}[t]
\centering

\includegraphics[width=0.7\columnwidth]{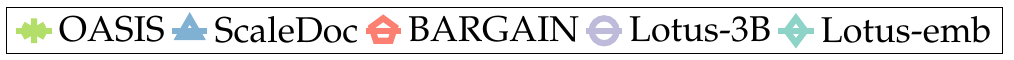}

\subfigure[\textsc{PubMed}]{%
    \includegraphics[width=0.33\columnwidth]{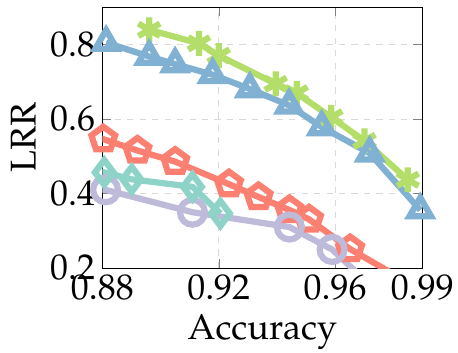}
    \label{fig:exp_sim_pubmed}
}%
\subfigure[\textsc{BigPatent}]{%
    \includegraphics[width=0.33\columnwidth]{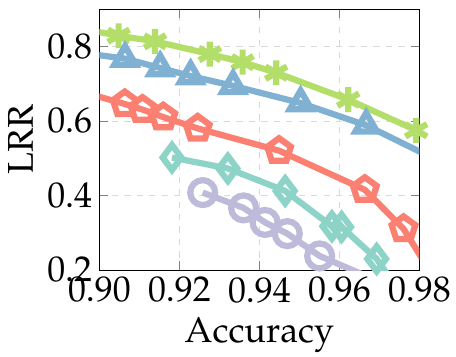}
    \label{fig:exp_sim_bigpatent}
}%
\subfigure[\textsc{GovReport}]{%
    \includegraphics[width=0.33\columnwidth]{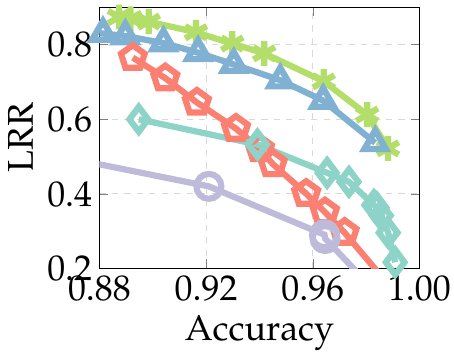}
    \label{fig:exp_sim_govreport}
}%

\caption{Accuracy--LRR trade-off on three datasets.}
\label{fig:exp1_sim}
\end{figure}

\begin{table}[t]
\centering
\caption{Comparison of monetary cost, online latency, and final accuracy with TaskCascade under target accuracy 0.9.}
\label{tab:cost_latency}
\scriptsize
\setlength{\tabcolsep}{3pt}
\begin{tabular}{lcccccccccc}
\toprule
\multirow{3}{*}{Dataset}
& \multicolumn{4}{c}{Monetary Cost}
& \multicolumn{4}{c}{Online Latency}
& \multicolumn{2}{c}{Acc.} \\
\cmidrule(lr){2-5} \cmidrule(lr){6-9} \cmidrule(lr){10-11}
& \multirow{2}{*}{\OASIS}
& \multicolumn{3}{c}{TC}
& \multirow{2}{*}{\OASIS}
& \multicolumn{3}{c}{TC}
& \multirow{2}{*}{\OASIS}
& \multirow{2}{*}{TC} \\
\cmidrule(lr){3-5} \cmidrule(lr){7-9}
&
& Opt. & Casc. & Total
&
& Opt. & Casc. & Total
& & \\
\midrule
\textsc{PubMed}
& \textbf{\$2.74} & \$3.53 & \$4.64 & \$8.17
& \textbf{376s} & 561s & 712s & 1273s
& \textbf{0.93} & 0.92 \\

\textsc{BigPatent}
& \textbf{\$0.82} & \$0.65 & \$0.97 & \$1.52
& \textbf{272s} & 144s & 291s & 435s
& \textbf{0.93} & 0.92 \\

\textsc{GovReport}
& \textbf{\$4.32} & \$5.31 & \$6.77 & \$12.08
& \textbf{602s} & 416s & 844s & 1260s
& \textbf{0.93} & 0.91 \\
\bottomrule
\end{tabular}
\end{table}

\section{Experimental Studies}
\label{sec:exp}

In this section, we present our experimental settings in \cref{sec:exp_setup} and report comprehensive evaluation in the following facets:  
\ding{172} Compare \OASIS with baselines on Accuracy and LLM reduction trade-off, online latency, monetary cost, and performance under workloads with different selectivities (\cref{sec:overall_eva});
\ding{173} Assess the impact and building cost of document NSG in offline phase  (\cref{sec:offline_eva});
\ding{174} Investigate the effectiveness of the GAT proxy model, the active learning strategy and the adaptive calibration sampling in online phase (\cref{sec:online_eva}).
\subsection{Experiment Setup}
\label{sec:exp_setup}

\stitle{Datasets and Workloads.}
We use three datasets that have been widely adopted in prior work \cite{scaledoc, taskcascade}: \textsc{PubMed}, \textsc{BigPatent}, and \textsc{GovReport}. Each dataset contains 10,000 documents, and the dataset statistics and example query are reported in Tab. \ref{tab:dataset_statistics}. For each dataset, we generate 10 semantic filter predicates, covering a wide range of filtering selectivities and query types, such as topic judgment and fine-grained content understanding. For each query, we prompt GPT-4o to obtain binary groundtruth labels.

\stitle{Baselines.}
We compare against five baselines. (1) \textbf{ScaleDoc.} \cite{scaledoc} It adopts an MLP-based contrastive learning method to obtain proxy scores, and uses a novel calibration workflow to select the ambiguous interval. (2) \textbf{BARGAIN.} \cite{bargain} It uses a much cheaper LLM to obtain proxy scores, and employs a progressive sampling algorithm to select the ambiguous interval while providing strong statistical accuracy guarantees. (3) \textbf{Lotus-3B.} \cite{lotus} It also uses a cheaper LLM to obtain proxy scores, and learns a threshold interval $[l, r]$ from sampled data as the ambiguous region. (4) \textbf{Lotus-emb.} \cite{lotus} It uses the same threshold-selection algorithm as Lotus-3B, but computes embedding similarity of each document and query as the proxy score. (5) \textbf{TaskCascade (TC).} \cite{taskcascade} It proposes task cascade which decomposes the query into easier sub-queries and combine it with document chunking and model cascade.

\begin{figure}[t]
    \centering
    \includegraphics[scale=0.63]{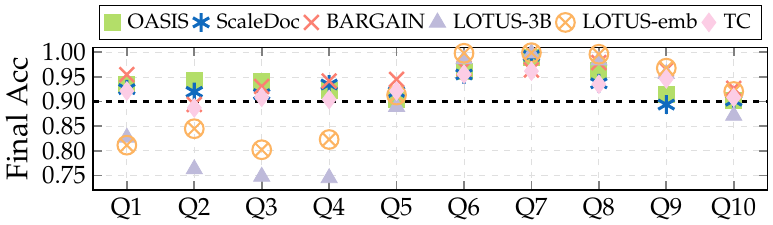}
    \caption{Final achieved accuracy on \textsc{PubMed}.}
    \label{fig:accvs}
\end{figure}

\begin{figure}[t]
\centering

\includegraphics[width=0.7\columnwidth]{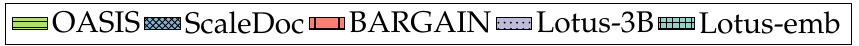}

\subfigure[\textsc{PubMed}]{%
    \includegraphics[width=0.32\columnwidth]{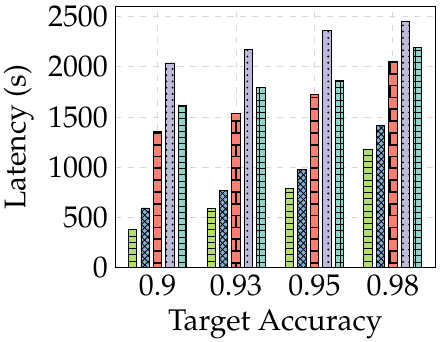}
    \label{fig:exp2_sim_pubmed}
}%
\subfigure[\textsc{BigPatent}]{%
    \includegraphics[width=0.32\columnwidth]{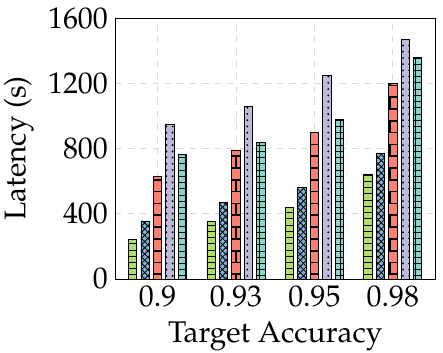}
    \label{fig:exp2_sim_bigpatent}
}%
\subfigure[\textsc{GovReport}]{%
    \includegraphics[width=0.32\columnwidth]{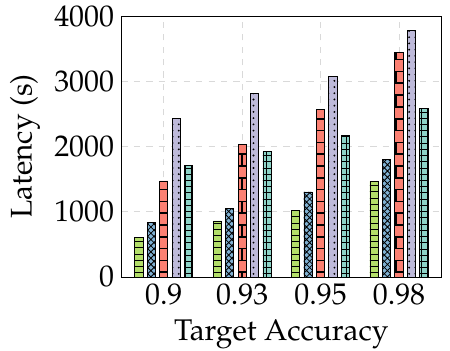}
    \label{fig:exp2_sim_govreport}
}%

\caption{Query latency under different accuracy targets.}
\label{fig:exp2_sim2}
\end{figure}

\begin{figure}[t]
\centering

\includegraphics[width=0.7\columnwidth]{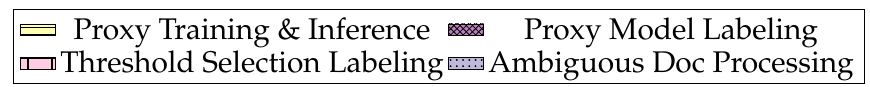}

\subfigure[\textsc{PubMed}]{%
    \includegraphics[width=0.47\columnwidth]{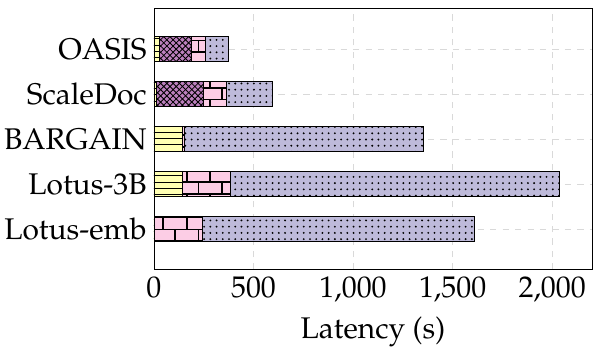}   
    \label{fig:exp3_sim_pubmed}
}%
\subfigure[\textsc{BigPatent}]{%
    \includegraphics[width=0.49\columnwidth]{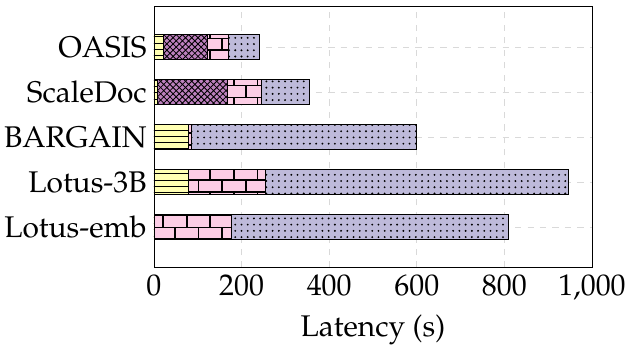}
    \label{fig:exp3_sim_govreport}
}%

\caption{Breakdown analysis of query latency with target accuracy 0.9.}
\label{fig:exp3_sim3}
\end{figure}

\stitle{Evaluation Metrics.}
We use the Accuracy--LLM Reduction Rate (LRR) curve to characterize the trade-off between query accuracy and LLM cost. \add{The Accuracy refers to the final achieved accuracy,} and the LLM Reduction Rate (LRR) is defined as the fraction of documents that do not require oracle LLM processing among all documents. The documents that require LLM oracle processing include the sampled documents in proxy training phase and threshold selection phase, and the documents falling into the ambiguous interval $[l,r]$. We also use the monetary cost of LLM as a metric according to the OpenAI API pricing \cite{openai}. For query efficiency, we use offline latency and online latency as evaluation metrics. To evaluate the effectiveness of the proxy model, we use AUC metric to measure its separation capability. A higher AUC indicates a clearer separation between positive and negative documents, which is more favorable for narrowing the subsequent thresholds.

\stitle{Implementation Details.}
For ScaleDoc, we use 10\% of the documents as the training set for the proxy model and 5\% of the documents as the calibration set, following the optimal configuration reported in the original paper. For Lotus, we use 10\% of the documents for threshold selection. For all embedding-based methods, we use Qwen3-Embedding-8B as the encoder. For the proxy models in BARGAIN and Lotus, we use Llama-3.2-3B. For the LLM oracle, we use GPT-4o through API calls with a rate limit of 150K tokens per minute. For TaskCascade, we use the same configuration for all LLMs in \cite{taskcascade}. Unless otherwise specified, we use the following default hyperparameters: $\add{m=20}, k=200, r=50, L_{num}=2,\lambda_{sup} = 0.2, \lambda_{reg} = 0.1, B_{ini}=400, \Delta B=100, B_{total}=1000, \add{\alpha_{exp}=0.35}, \tau_{val}=0.8, \eta_1=0.08, \eta_2=0.08$, $\eta_3=0.09, |\mathcal{R}|=200,000$, $a_{ini}=2$, and $\Delta a=0.5$. All experiments are conducted on a server with a RTX 5090 GPU and a 25-core Intel(R) Xeon(R) Platinum 8470Q CPU. \add{We run each method five times and report the average results.}

\begin{figure}[t]
\centering
\includegraphics[width=0.7\columnwidth]{exp/exp-1/legend-exp1.pdf}

\subfigure[\textsc{PubMed}-Low]{%
    \includegraphics[width=0.32\columnwidth]{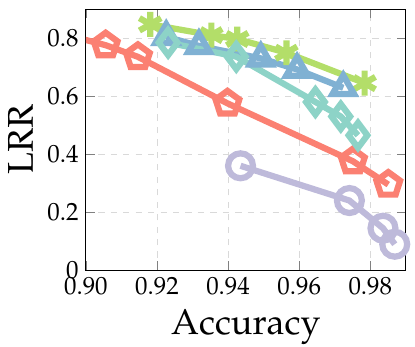}
    \label{fig:exps_sim_pubmedlow}
}%
\subfigure[\textsc{PubMed}-High]{%
    \includegraphics[width=0.32\columnwidth]{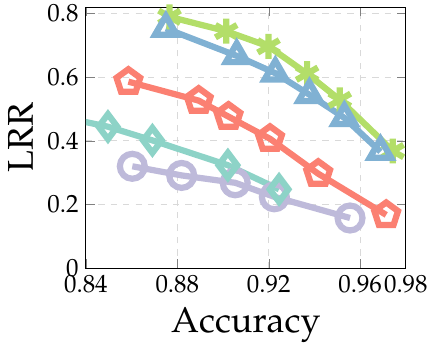}
    \label{fig:exps_sim_pubmedhigh}
}%
\subfigure[\textsc{BigPatent}-Low]{
    \includegraphics[width=0.32\columnwidth]{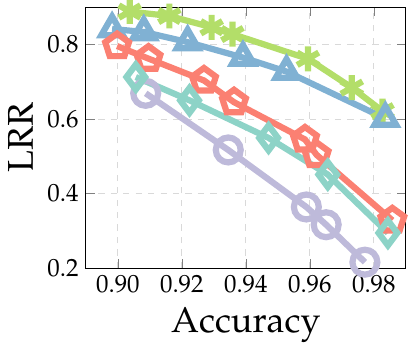}
    \label{fig:exps_sim_bigpatentlow}
}
\subfigure[\textsc{BigPatent}-High]{%
    \includegraphics[width=0.32\columnwidth]{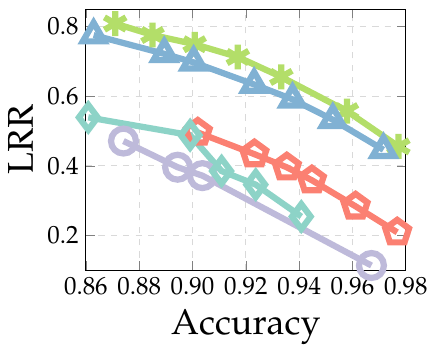}  
    \label{fig:exps_sim_bigpatenthigh}
}%
\subfigure[\textsc{GovReport}-Low]{%
    \includegraphics[width=0.32\columnwidth]{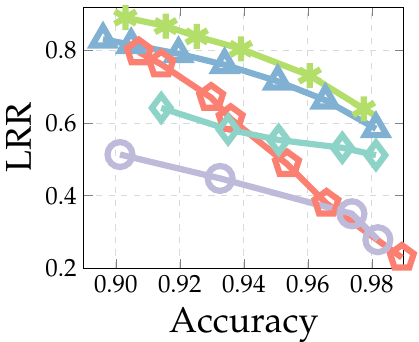}
    \label{fig:exps_sim_govreportlow}
}%
\subfigure[\textsc{GovReport}-High]{%
    \includegraphics[width=0.32\columnwidth]{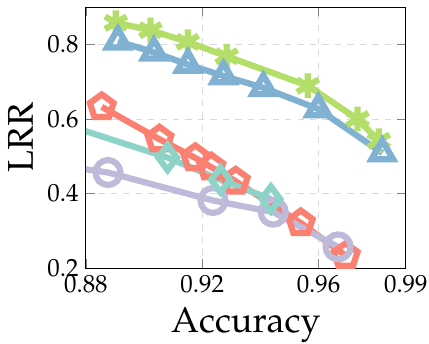}    
    \label{fig:exps_sim_govreporthigh}
}%

\caption{Accuracy-LRR under workloads with different selectivities.}
\label{fig:expselect_sim}
\end{figure}

\subsection{Overall Evaluation}
\label{sec:overall_eva}
\stitle{Accuracy-LRR and Cost Comparison.} 
We first compare the LLM Reduction Rate of each method under different accuracy levels. Fig. \ref{fig:exp1_sim} presents the Accuracy--LRR curves of \OASIS and other baselines on the three datasets. Note that we exclude TC from this comparison, because it uses multiple LLMs rather than a single LLM oracle. Therefore, we compare with it in terms of the monetary cost of all LLMs, as shown in Tab. \ref{tab:cost_latency}.

The results show that our method significantly outperforms all baselines across all datasets under the same accuracy. \add{Specifically, at a final accuracy of 0.9, \OASIS achieves 85\% LRR on average, requiring only 15\% LLM-processed documents, while Lotus-3B, Lotus-emb, BARGAIN, and ScaleDoc require 60\%, 48\%, 38\%, 23\%, respectively, corresponding to a 35\%–75\% reduction in LLM calls relative to these baselines. Compared with the strongest baseline ScaleDoc, \OASIS outperforms it on 29 out of 30 queries.} This is because \OASIS adopts a GNN as the proxy model to obtain more separable proxy scores, and further use online adaptive active learning and calibration set selection to reduce LLM cost while maintaining high accuracy. Lotus-3B and BARGAIN use the output logits of a 3B LLM as proxy scores. Since small models lack sufficient reasoning capability for complex predicates, the generated proxy scores are often unreliable, causing a large number of documents to be processed by the LLM oracle during subsequent threshold selection. Lotus-emb uses the embedding similarity between the document and the query as the proxy score, which produces better proxy scores than the 3B model. However, Lotus relies on fixed-size sampling for threshold selection, and the learned thresholds may lead to a noticeable gap between the achieved accuracy and the target accuracy, resulting in higher LLM cost. ScaleDoc uses a simple MLP to learn proxy scores, which ignores the clustering characteristics of similar documents. Moreover, it uses a fixed number of sampled documents for both proxy learning and distribution reconstruction. This lack of flexibility can amplify the LLM cost. Compared with TC, \OASIS reduces the monetary cost by 1.85$\times$--2.98$\times$. This is because \OASIS can effectively control the LLM budget through adaptive learning and sampling, whereas the small LLMs used by TC to obtain proxy scores are also less reliable, and constructing the task cascade and model cascade incurs high cost.

\add{Fig. \ref{fig:accvs} compares the final achieved accuracy by different methods under a target accuracy of 0.9 (we report only the results on \textsc{PubMed} due to space limit). The results show that \OASIS is the only method that consistently meets the target accuracy, demonstrating the effectiveness and reliability of its progressive calibration.}

\stitle{Latency Comparison and Breakdown Analysis.}
Fig. \ref{fig:exp2_sim2} and Tab. \ref{tab:cost_latency} report the online latency under different target accuracies. The results show that \OASIS achieves lower latency than the baselines across all target accuracy settings. Specifically, at 0.9 target accuracy, the online latency of \OASIS is reduced by 1.75$\times$--5.42$\times$ on average compared with all baselines. This is because the GNN-based proxy model with the adaptive learning and sampling strategies in \OASIS reduces both the number of sampled documents and final LLM judgment documents, thereby reducing the online latency.

\begin{table}[t]
\centering
\footnotesize
\caption{Accuracy, F1, and LRR on high- and low-selectivity
groups with target acc = 0.9.}

\setlength{\tabcolsep}{3pt}
\begin{tabular}{cc|ccc|ccc}
\hline
\multirow{2}{*}{Dataset}
& \multirow{2}{*}{Method}
& \multicolumn{3}{c|}{High}
& \multicolumn{3}{c}{Low} \\
& & Acc. & F1 & LRR & Acc. & F1 & LRR \\
\hline

\multirow{2}{*}{\textsc{PubMed}}
& OASIS
& 0.9218 & 0.8685 & 0.6906 
& 0.9388 & 0.9191 & 0.8296 \\
& ScaleDoc
& 0.9123 & 0.8652 & 0.6524
& 0.9331 & 0.8676 & 0.7597 \\
\hline

\multirow{2}{*}{\textsc{BigPatent}}
& OASIS
& 0.9299 & 0.9024 & 0.6663
& 0.9403 & 0.7917 & 0.8363 \\
& ScaleDoc
& 0.9231 & 0.8796 & 0.6136
& 0.9283 & 0.7256 & 0.7827 \\
\hline

\multirow{2}{*}{\textsc{GovReport}}
& OASIS
& 0.9208 & 0.8349 & 0.8004
& 0.9324 & 0.7382 & 0.8270 \\
& ScaleDoc
& 0.9151 & 0.8043 & 0.7260
& 0.9196 & 0.5908 & 0.8014 \\
\hline

\end{tabular}
\label{tab:selectivity_f1}
\end{table}

\begin{figure}[t]
\centering

\includegraphics[width=0.7\columnwidth]{exp/exp-1/legend-exp1.pdf}

\subfigure[GPT-4o-mini]{%
    \includegraphics[width=0.33\columnwidth]{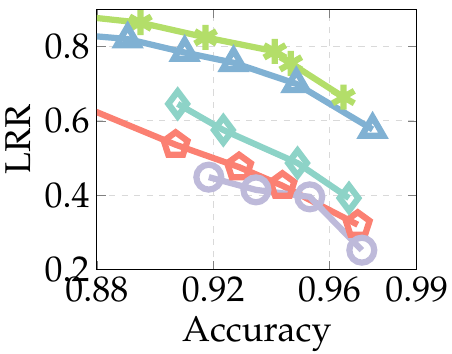}
    \label{fig:exp_sim_pubmed}
}%
\subfigure[Deepseek-v4-flash]{%
    \includegraphics[width=0.33\columnwidth]{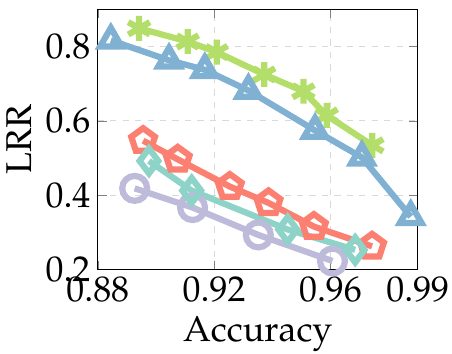}
    \label{fig:exp_sim_bigpatent}
}%
\subfigure[Qwen3.5-35B]{%
    \includegraphics[width=0.33\columnwidth]{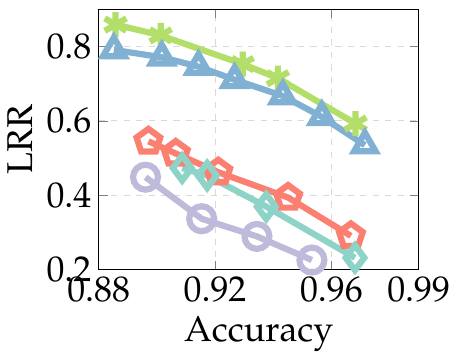}
    \label{fig:exp_sim_govreport}
}%

\caption{Generalization on more oracle LLMs.}
\label{fig:exp_oracle}
\end{figure}

\begin{table*}[t]
\centering
\footnotesize
\setlength{\tabcolsep}{3.5pt}
\caption{The AUC value of proxy scores by different proxy models on three datasets (The best result is bolded).}
\label{tab:proxy-model-auc}
\begin{tabular}{lcccccccccccc}
\toprule
\multirow{2}{*}{Query} 
& \multicolumn{4}{c}{\textsc{PubMed}} 
& \multicolumn{4}{c}{\textsc{BigPatent}} 
& \multicolumn{4}{c}{\textsc{GovReport}} \\
\cmidrule(lr){2-5} \cmidrule(lr){6-9} \cmidrule(lr){10-13}
& \OASIS & ScaleDoc & Llama-3B & Embedding
& \OASIS & ScaleDoc & Llama-3B & Embedding
& \OASIS & ScaleDoc & Llama-3B & Embedding \\
\midrule
Q1  & \textbf{0.8814} & 0.8626 & 0.7222 & 0.6755 & \textbf{0.9821} & 0.9795 & 0.8159 & 0.8754 & \textbf{0.9137} & 0.9031 & 0.7420 & 0.8628 \\
Q2  & \textbf{0.8157} & 0.7822 & 0.6988 & 0.7304 & \textbf{0.9790} & 0.9751 & 0.8719 & 0.9202 & \textbf{0.8902} & 0.8632 & 0.6820 & 0.8288 \\
Q3  & \textbf{0.8863} & 0.8668 & 0.5803 & 0.6546 & \textbf{0.9804} & 0.9731 & 0.8565 & 0.8331 & \textbf{0.8787} & 0.8673 & 0.6658 & 0.8101 \\
Q4  & \textbf{0.8819} & 0.8778 & 0.6293 & 0.7050 & \textbf{0.9738} & 0.9627 & 0.6984 & 0.8925 & \textbf{0.8360} & 0.8302 & 0.6642 & 0.7469 \\
Q5  & \textbf{0.9218} & 0.9060 & 0.6766 & 0.7846 & \textbf{0.8318} & 0.8282 & 0.7082 & 0.7053 & \textbf{0.9452} & 0.9394 & 0.7788 & 0.8749 \\
Q6  & \textbf{0.9783} & 0.9778 & 0.7286 & 0.9406 & \textbf{0.8926} & 0.8812 & 0.7239 & 0.7637 & \textbf{0.9711} & 0.9699 & 0.7194 & 0.8829 \\
Q7  & 0.9914 & \textbf{0.9968} & 0.8193 & 0.9871 & \textbf{0.8727} & 0.8579 & 0.6692 & 0.6992 & \textbf{0.9561} & 0.9354 & 0.6977 & 0.9158 \\
Q8  & \textbf{0.9600} & 0.9498 & 0.7826 & 0.8802 & \textbf{0.9244} & 0.9121 & 0.6548 & 0.8468 & \textbf{0.8960} & 0.8832 & 0.6882 & 0.8596 \\
Q9  & \textbf{0.8669} & 0.8546 & 0.5382 & 0.6950 & \textbf{0.8422} & 0.8225 & 0.6123 & 0.6636 & \textbf{0.9164} & 0.8882 & 0.7168 & 0.8644 \\
Q10 & \textbf{0.7494} & 0.7471 & 0.5129 & 0.6198 & \textbf{0.9173} & 0.9038 & 0.6911 & 0.8842 & \textbf{0.9142} & 0.8988 & 0.7027 & 0.8209 \\
\bottomrule
\end{tabular}
\end{table*}

Fig. \ref{fig:exp3_sim3} presents the latency breakdown of each stage for the baselines. The results show that BARGAIN and Lotus-3B use a small LLM as the proxy model, which leads to long inference time and makes proxy inference a significant cost. In contrast, the other baselines are based on embedding similarity. Since the embeddings of all documents can be constructed offline, the online phase only needs to encode the query into the query embedding and compute similarity scores, making this overhead negligible. Since \OASIS adopts multi-round active learning, its proxy training and inference time is slightly higher than that of ScaleDoc. However, this additional computation produces better proxy scores, resulting in fewer documents falling into the ambiguous interval. The reduced LLM processing time can fully compensate for the additional proxy training cost. Moreover, due to the early-stopping and adaptive sampling mechanisms in \OASIS, both the number of documents and the time cost for proxy model training labeling and threshold selection samples labeling are lower than those of ScaleDoc.

\begin{figure}[t]
    \centering
    \begin{minipage}[c]{0.45\columnwidth}
        \centering
        \includegraphics[width=\linewidth]{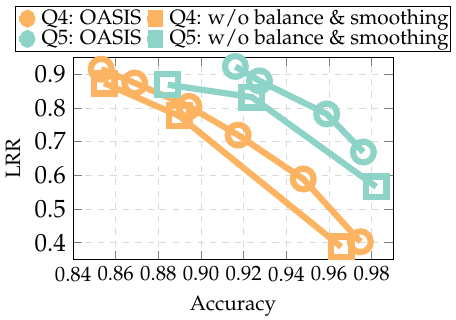}
        \captionof{figure}{Effect of balance and smoothing.}
        \label{fig:bce-ablation}
    \end{minipage}%
    \hfill
    \begin{minipage}[c]{0.53\columnwidth}
        \centering
        \footnotesize
        \setlength{\tabcolsep}{1.0pt}
        \renewcommand{\arraystretch}{1.12}

        \captionof{table}{Offline construction time (s).}
        \label{tab:offline-construction-time}

        \begin{tabular}{@{}lrrr@{}}
            \toprule
            \multirow{2}{*}{Dataset}
            & \multicolumn{1}{c}{Edge}
            & \multicolumn{1}{c}{KNN}
            & \multirow{2}{*}{Embed.} \\
            & \multicolumn{1}{c}{Refine}
            & \multicolumn{1}{c}{Construct}
            & \\
            \midrule
            \textsc{PubMed}    & 2.61 & 1.65 & 333.75 \\
            \textsc{BigPatent} & 2.33 & 1.47 & 95.69 \\
            \textsc{GovReport} & 2.45 & 1.54 & 445.69 \\
            \bottomrule
        \end{tabular}
    \end{minipage}
\end{figure}

\stitle{Performance under Different Selectivity Workloads.}
We further study how queries with different selectivities affect performance. For the 10 queries in each dataset, we define low-selectivity queries as those where fewer than 30\% of documents pass the filter, and high-selectivity queries as those where more than 30\% of documents pass the filter. Fig. \ref{fig:expselect_sim} reports the average results of low-selectivity and high-selectivity queries on each dataset.

The experimental results show that \OASIS outperforms all baselines under both types of workloads across all accuracy settings, demonstrating that \OASIS can adapt to workloads with diverse characteristics. This is because the online adaptive active learning strategy and calibration sampling strategy can adaptively select different numbers of sampled documents according to query characteristics, thereby reducing LLM cost. Notably, high-selectivity queries are more challenging than low-selectivity queries. For all methods, the LRR of high-selectivity queries is 10\%--20\% lower than that of low-selectivity queries under the same accuracy. This is because the numbers of positive and negative documents are more balanced, causing their score distributions to overlap more substantially and making threshold selection more difficult.

\add{Tab. \ref{tab:selectivity_f1} compares the F1 score, accuracy, and LRR of \OASIS and ScaleDoc at a target accuracy of 0.9. The results show that \OASIS consistently outperforms ScaleDoc across all three metrics under the same interface, with particularly substantial F1 improvements in high- and low-selectivity scenarios.}

\stitle{Generalization on More Oracle LLMs.} \add{We further evaluate \OASIS using GPT-4o-mini, DeepSeek-v4-flash, and Qwen3.5-35B as alternative oracle models, with the results presented in Fig. \ref{fig:exp_oracle}. \OASIS consistently maintains its performance advantages across these models, demonstrating strong generalizability to LLMs with varying model sizes and capabilities.}

\subsection{Analysis of Offline Phase}
\label{sec:offline_eva}

\stitle{Efficiency on Offline Phase.} 
We investigate the time overhead of constructing the document NSG in the offline stage. Tab. \ref{tab:offline-construction-time} reports the time cost of each stage in constructing the document NSG on the three datasets. The experimental results show that LLM encoding dominates the overall offline time. In contrast, the time for constructing the document NSG is relatively very small. Although the NSG construction contains a quadratic term, this is a one-time, query-agnostic offline preprocessing cost. For substantially larger collections, NSG construction can also be accelerated using GPU-based similarity computation and parallel processing without compromising scalability.

\begin{figure}[t]
\centering

\includegraphics[width=0.37\columnwidth]{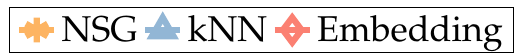}

\subfigure[\textsc{PubMed}]{%
    \includegraphics[width=0.32\columnwidth]{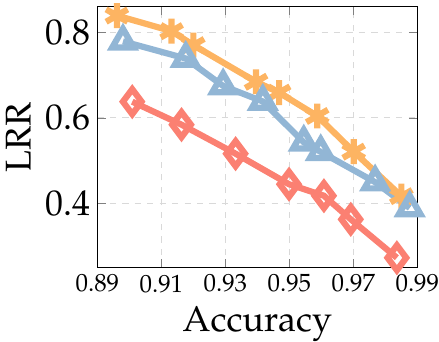}
    \label{fig:exp5_sim_pubmed}
}%
\subfigure[\textsc{BigPatent}]{%
    \includegraphics[width=0.32\columnwidth]{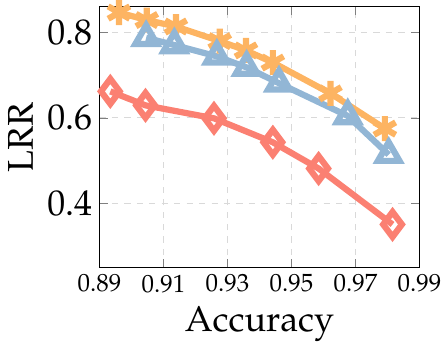}
    \label{fig:exp5_sim_bigpatent}
}%
\subfigure[\textsc{GovReport}]{ %
    \includegraphics[width=0.32\columnwidth]{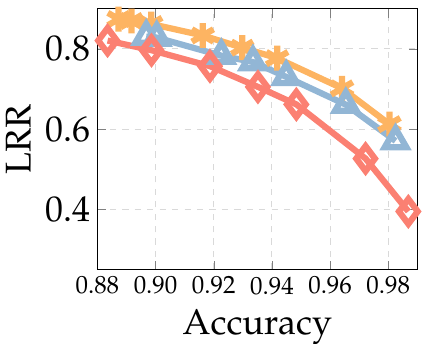}
    \label{fig:exp5_sim_govreport}
}%

\caption{Accuracy-LRR under different proxy structures.}
\label{fig:exp5_sim}
\end{figure}

\stitle{Effectiveness of Document NSG.}
We investigate the effectiveness of different proxy structures. The first uses the document NSG for GNN training. The second uses the kNN graph without edge refinement as the backbone, while keeping the GNN architecture and training procedure the same. The third uses the embedding similarity between the query and document as the proxy score. All three methods use the same threshold selection algorithm as \OASIS. Fig. \ref{fig:exp5_sim} presents the experimental results. The results show that NSG achieves better LLM reduction than kNN. This is because NSG is sparser than kNN, while removing redundant edges and preserving important neighboring relationships. As a result, it not only improves the efficiency of GNN training, but also leads to better semi-supervised learning performance. The method using only embedding similarity performs worse than the learning-based methods, because the benefit of learning outweighs the overhead of sampling training examples, reducing more ambiguous documents in the threshold selection phase.

\subsection{Analysis of Online Phase}
\label{sec:online_eva}
\stitle{Effectiveness of Proxy model.}
We compare the quality of proxy scores produced by different proxy models. For proxy scores, we use the AUC metric to measure the effectiveness of the proxy model. If the proxy model separates the scores of positive and negative documents more clearly, it achieves a higher AUC, which is more favorable for the subsequent threshold selection stage. In this case, fewer documents fall into the selected uncertain interval which need to be processed by the LLM oracle. For \OASIS, we remove the active learning phase and use fixed 10\% of the documents for training, the same setting as ScaleDoc. Llama-3B and Embedding denote the methods that use the output logits of Llama-3B and the query-document embedding similarity as proxy scores, respectively. The results are shown in Tab. \ref{tab:proxy-model-auc}. The results show that the GAT-based proxy model in \OASIS achieves a significant advantage, obtaining the highest AUC score on 29 out of 30 queries. This is because GAT facilitates the propagation of anchor labeling signals and learns the clustering characteristics of neighboring documents, thereby outperforming MLP. In contrast, Llama-3B is limited by its small model size and has insufficient reasoning capability, resulting in less reliable scores. Although the embeddings generated by the 8B model contain richer semantic information than those produced by the 3B model, they still underperform the sampling and learning based method, where a small number of sampled documents can bring larger benefits.

\begin{figure}[t]
\centering

\subfigure[\textsc{PubMed}]{%
    \includegraphics[width=0.3\columnwidth]{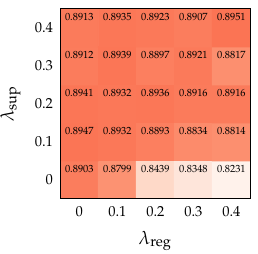}
    \label{fig:exp8_sim_pubmed}
}%
\subfigure[\textsc{BigPatent}]{%
    \includegraphics[width=0.3\columnwidth]{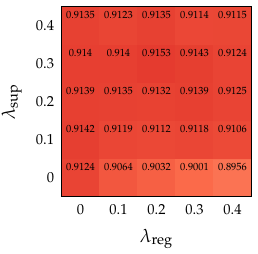}
    \label{fig:exp8_sim_bigpatent}
}%
\subfigure[\textsc{GovReport}]{%
    \includegraphics[width=0.3\columnwidth]{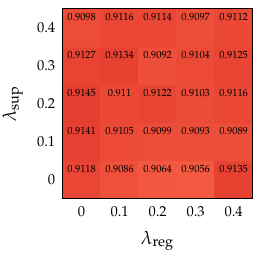}
    \label{fig:exp8_sim_govreport}
}%
\includegraphics[width=0.06\linewidth]{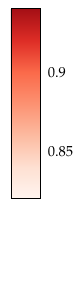}
\caption{Hyper-parameter effects on GNN learning phase.}
\label{fig:exp6_sim}
\end{figure}

\begin{figure}[t]
\centering

\includegraphics[width=0.7\columnwidth]{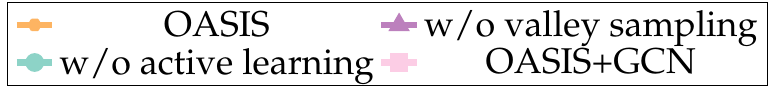}

\subfigure[\textsc{PubMed}]{%
    \includegraphics[width=0.32\columnwidth]{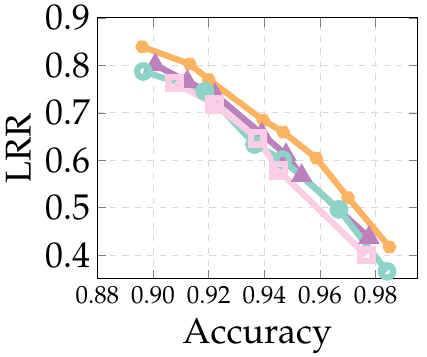}
    \label{fig:exp6_sim_pubmed_new}
}%
\subfigure[\textsc{BigPatent}]{  %
    \includegraphics[width=0.32\columnwidth]{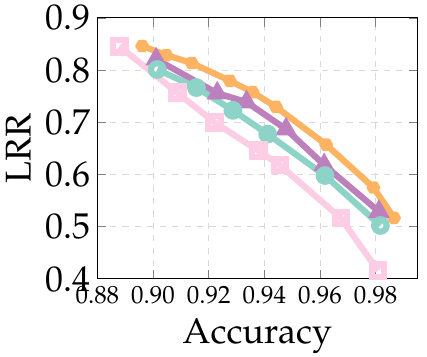}
    \label{fig:exp6_sim_bigpatent_new}
}%
\subfigure[\textsc{GovReport}]{   %
    \includegraphics[width=0.32\columnwidth]{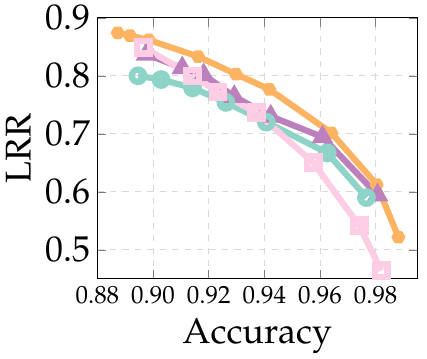}
    \label{fig:exp6_sim_govreport_new}
}%

\caption{Ablation study of active learning and GNN.}
\label{fig:expactive_sim_new}
\end{figure}

\begin{figure}[t]
\centering

\includegraphics[width=0.7\columnwidth]{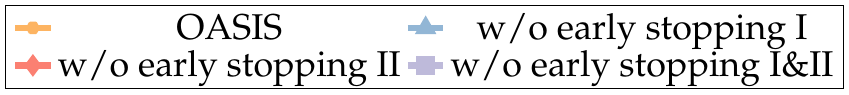}

\subfigure[\textsc{PubMed}]{%
    \includegraphics[width=0.32\columnwidth]{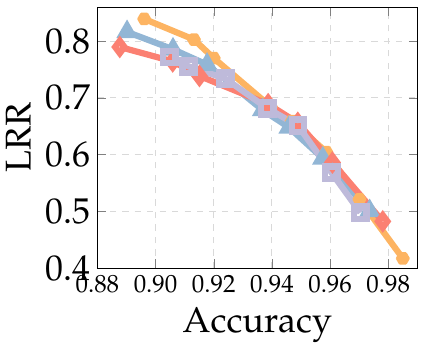}
    \label{fig:exp6_sim_pubmed}
}%
\subfigure[\textsc{BigPatent}]{  %
    \includegraphics[width=0.32\columnwidth]{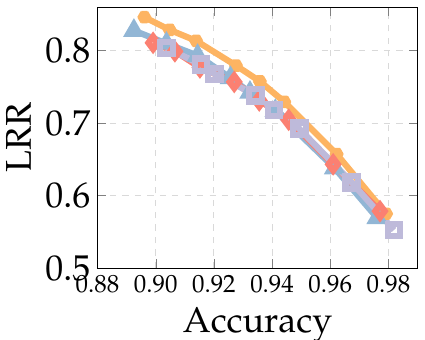}
    \label{fig:exp6_sim_bigpatent}
}%
\subfigure[\textsc{GovReport}]{   %
    \includegraphics[width=0.32\columnwidth]{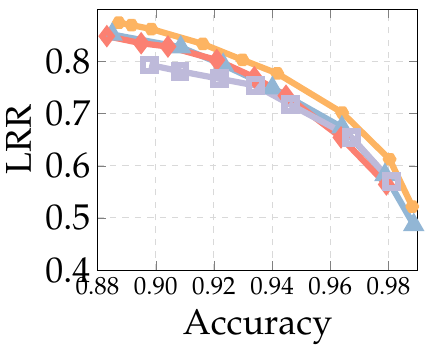}
    \label{fig:exp6_sim_govreport}
}%

\caption{Ablation study of early stopping.}
\label{fig:expactive_sim}
\end{figure}

\stitle{Hyper-parameter Effect on GNN learning.}
We investigate the effects of two parameters, $\lambda_{sup}$ and $\lambda_{reg}$ in GAT training. We vary $\lambda_{sup}$ and $\lambda_{reg}$ over $[0, 0.1, 0.2, 0.3, 0.4] \times [0, 0.1, 0.2, 0.3, 0.4]$, use 10\% of the documents as training samples, and remove active learning. We then compute the AUC achieved by the trained proxy model. The experimental results are shown in the heatmaps in Fig. \ref{fig:exp6_sim}. The results show that settings with $\lambda_{sup} > 0$ generally outperform $\lambda_{sup}=0$. This is because the supervised contrastive loss can separate nodes with different labels in the latent space, leading to higher AUC of the proxy scores. In contrast, varying $\lambda_{reg}$ does not show a significant impact on AUC. This is because the smoothness loss mainly smooths the score distribution, making the subsequent distribution reconstruction more reliable; therefore, its impact on AUC is weaker than that of the BCE loss and the supervised contrastive loss. Based on the observations over the three datasets, we set $\lambda_{sup}=0.2$ and $\lambda_{reg}=0.1$ by default, which achieves near-optimal performance across all datasets.

\stitle{Effect of Balance and Smoothing on Skewed Cases.} \add{We investigate the effects of label balancing and smoothing in BCE under skewed label distributions (i.e., queries with selectivity below 5\% on \textsc{PubMed}). As shown in Fig. \ref{fig:bce-ablation}, both techniques improve robustness to label imbalance and enhance learning effectiveness, leading to substantial improvements in LRR.}

\begin{figure}[t]
\centering

\includegraphics[width=0.4\columnwidth]{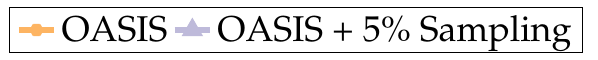}

\subfigure[\textsc{PubMed}]{%
    \includegraphics[width=0.32\columnwidth]{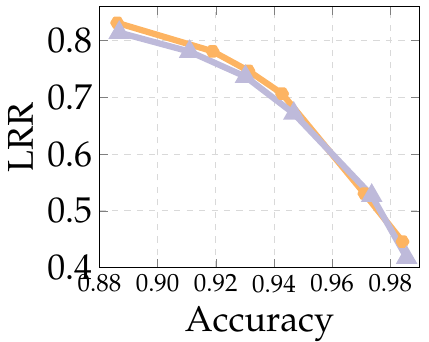}
    \label{fig:exp10_sim_pubmed}
}%
\subfigure[\textsc{BigPatent}]{%
    \includegraphics[width=0.32\columnwidth]{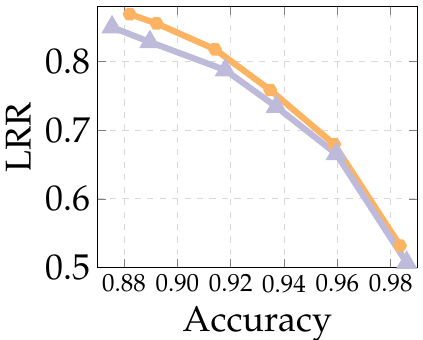}
    \label{fig:exp10_sim_bigpatent}
}%
\subfigure[\textsc{GovReport}]{%
    \includegraphics[width=0.32\columnwidth]{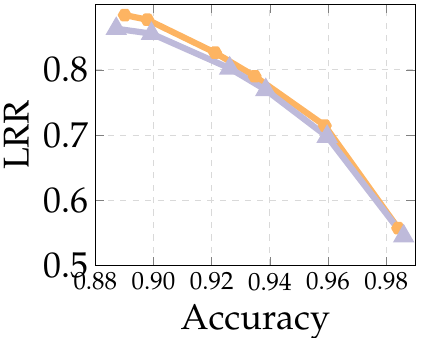}
    \label{fig:exp10_sim_govreport}
}%

\subfigure[\textsc{PubMed}-Q3]{%
    \includegraphics[width=0.32\columnwidth]{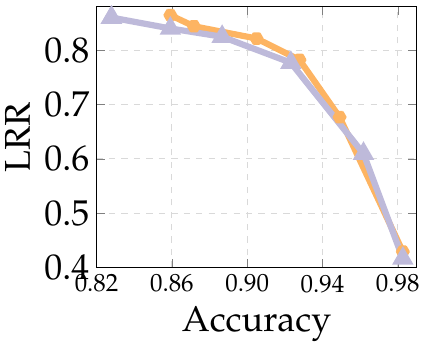}
    \label{fig:exp10_sim_pubmed2}
}%
\subfigure[\textsc{BigPatent}-Q1]{%
    \includegraphics[width=0.32\columnwidth]{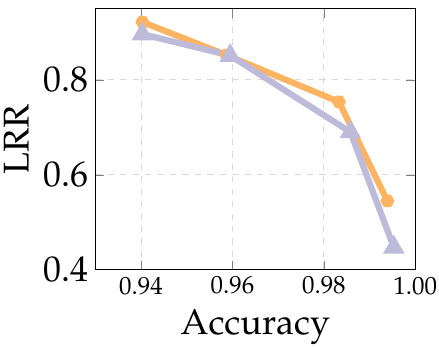}

    \label{fig:exp10_sim_bigpatent0}
}%
\subfigure[\textsc{GovReport}-Q6]{%
    \includegraphics[width=0.32\columnwidth]{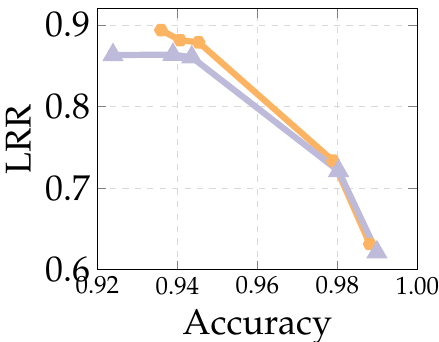}
    \label{fig:exp10_sim_govreport5}
}%

\caption{Ablation study of adaptive calibration set selection.}
\label{fig:exp10_sim}
\end{figure}

\stitle{Ablation Study of Active Learning and GAT.}
\add{We conduct ablation studies to investigate the role of each component in active learning. In addition to \OASIS, we implement six variants: (1) replacing valley sampling with full range sampling; (2) replacing active learning with one-time sampling and learning; (3) replacing GAT with GCN; (4) removing the early-stopping strategy I ; (5) removing the early-stopping strategy II; (6) removing both early-stopping strategies. The results in Fig. \ref{fig:expactive_sim} and Fig. \ref{fig:expactive_sim_new} show that the Accuracy--LRR curve of \OASIS is almost consistently above the curves of other variants. This indicates that the valley-based sampling and active learning can effectively leverage the score information from each training round to iteratively select the most informative samples for labeling and retraining, leading to better learning performance. The curves of the variants without early-stopping are generally lower than that of \OASIS, indicating that the early-stopping strategies can terminate training in time when the marginal benefit becomes limited, thereby saving the LLM labeling cost. GAT also outperforms GCN by considering the different importance of neighbors through attention-based weighting.}

\stitle{Effectiveness of adaptive calibration sampling.}
We conduct experiments to evaluate the effectiveness of adaptive calibration set sampling in the threshold-selection stage. For \OASIS, we compare it with a variant that removes adaptive sampling, i.e., fixed 5\% sampling within each bin, while keeping all other stages the same as \OASIS. The Accuracy--LRR curves are shown in Fig. \ref{fig:exp10_sim}. The results show that the Accuracy--LRR curves of \OASIS are generally above those of the variant on the three datasets. Specifically, at an actual accuracy of 0.9, the LLM Reduction Rate of the variant decreases by 2\% to 2.5\% on average. This demonstrates that the adaptive calibration set sampling in \OASIS can reduce certain overhead while maintaining comparable accuracy compared with fixed-size sampling. We further select one representative query from each dataset. \textsc{PubMed}-Q3 is a relatively difficult query, and thus its $\mathcal{V}_{ov}$ is larger, requiring more sampled documents. As a result, the curves are nearly overlapping. In contrast, \textsc{BigPatent}-Q1 and \textsc{GovReport}-Q6 are easier queries, with smaller $\mathcal{V}_{ov}$, where only a small number of sampled documents is sufficient for convergence. Therefore, the LLM reduction effect is more significant.


\section{Conclusion}
\label{sec:conclusion}
In this paper, we propose \OASIS, a framework towards online adaptive LLM-based semantic filtering. \OASIS combines NSG-based GAT active learning and adaptive threshold calibration, which improves proxy learning effectiveness and adapts LLM budgets to queries with different difficulties. We conduct extensive experiments to show that \OASIS significantly outperforms existing methods.

\balance
\bibliographystyle{ACM-Reference-Format}
\bibliography{sample-base}

\appendix

\end{document}